\documentclass[11pt]{article}
\usepackage[final]{acl}
\usepackage{xurl}
\usepackage{times}
\usepackage{latexsym}
\usepackage{amssymb}
\usepackage{pifont}
\usepackage[T1]{fontenc}
\usepackage[utf8]{inputenc}
\usepackage{microtype}
\usepackage{inconsolata}
\usepackage{graphicx}
\usepackage{booktabs}
\usepackage{lscape}
\usepackage{longtable}
\usepackage{array} 
\usepackage{ragged2e} 
\usepackage{xcolor}
\usepackage{tikz}
\usepackage[edges]{forest}
\definecolor{hidden-draw}{RGB}{20,68,106}
\definecolor{hidden-pink}{RGB}{255,245,247}
\usepackage{hyperref}   
\usepackage{fontawesome5} 
\usepackage{academicons}
\usepackage{svg}

\usepackage{amsmath,amsfonts,bm}
\usepackage[table]{xcolor} 
\usepackage{booktabs}
\usepackage{ragged2e}
\usepackage{hyperref} 

\def\eqref#1{equation~\ref{#1}}

\DeclareMathAlphabet{\mathsfit}{\encodingdefault}{\sfdefault}{m}{sl}
\SetMathAlphabet{\mathsfit}{bold}{\encodingdefault}{\sfdefault}{bx}{n}

\def\gC{{\mathcal{C}}}
\def\gD{{\mathcal{D}}}
\def\gE{{\mathcal{E}}}

\def\gI{{\mathcal{I}}}

\def\gM{{\mathcal{M}}}

\def\gP{{\mathcal{P}}}

\def\gT{{\mathcal{T}}}

\def\sI{{\mathbb{I}}}

\newcommand{\ghlink}[1]{\href{#1}{\textcolor{black}{\faGithub}}}
\newcommand{\faHuggingFace}{%
  \raisebox{-0.13em}{%
    \includegraphics[height=1em]{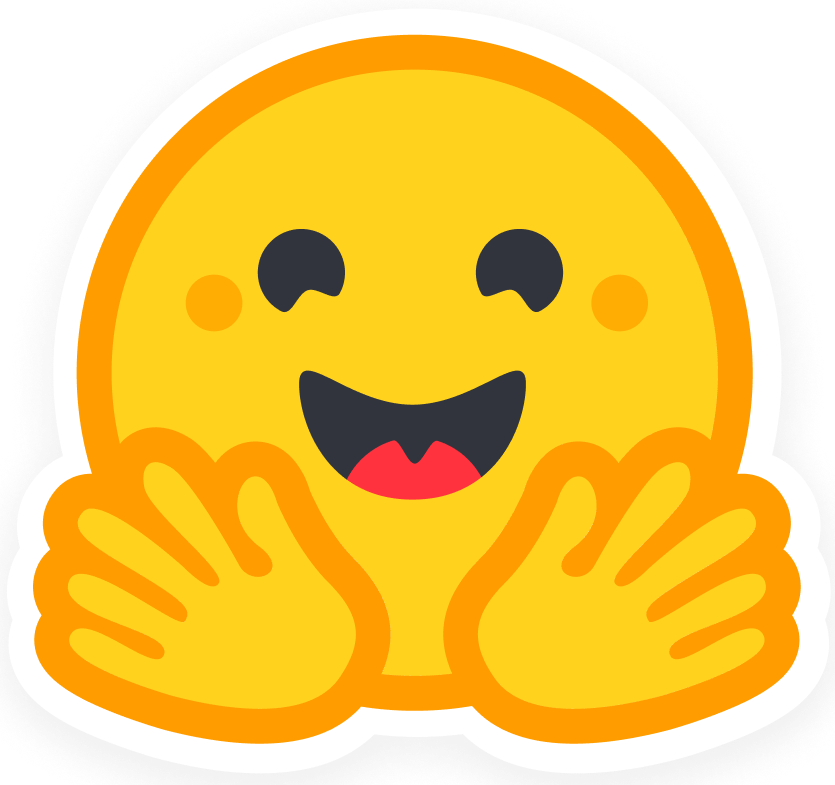}%
  }%
}
\newcommand{\hflink}[1]{\href{#1}{\textcolor{black}{\faHuggingFace}}}
\newcommand{\bloglink}[1]{\href{#1}{\color{blue}\faGlobe}}

\title{
\raisebox{-0.25\height}{\includegraphics[height=1.6em]{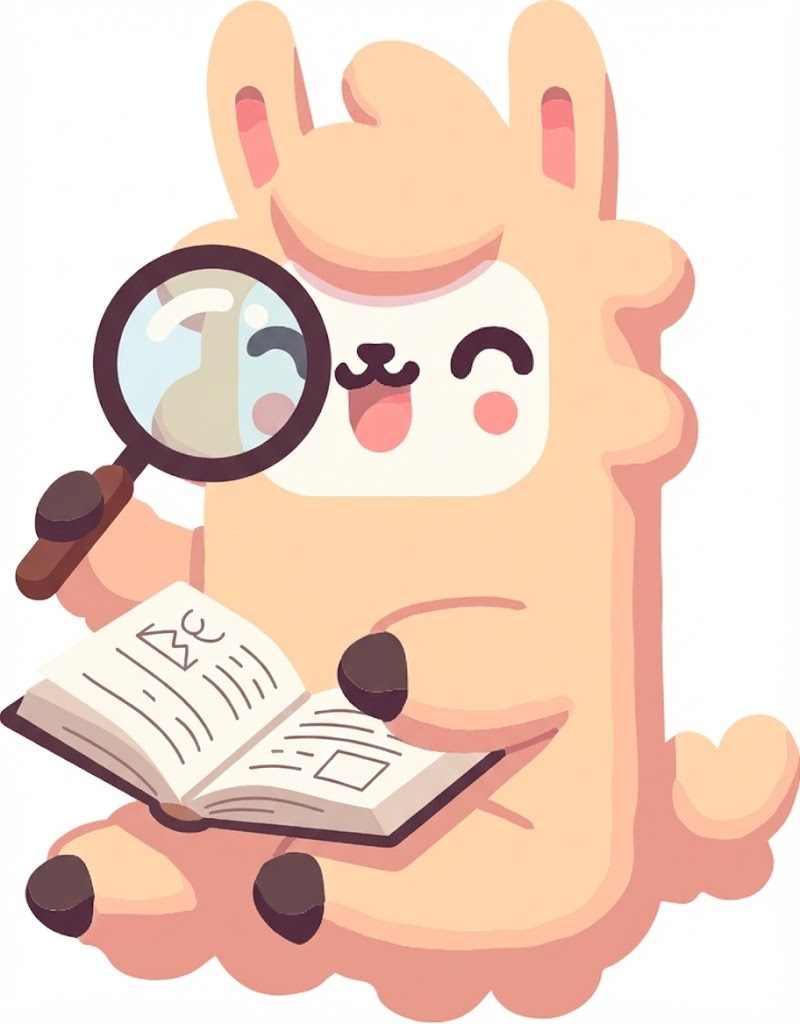}}%
  \hspace{-0.1em}
Advances and Frontiers of LLM-based Issue Resolution in Software Engineering: A Comprehensive Survey
}

\author{
    \textbf{Caihua Li}\textsuperscript{1}, 
    \textbf{Lianghong Guo}\textsuperscript{1}, 
    \textbf{Yanlin Wang}\textsuperscript{1}\thanks{Corresponding authors.}, 
    \textbf{Daya Guo}\textsuperscript{1},
    \textbf{Wei Tao}\textsuperscript{2}\footnotemark[1], 
    \textbf{Zhenyu Shan}\textsuperscript{3}, 
    \textbf{Mingwei Liu}\textsuperscript{1}, \\
    \textbf{Jiachi Chen}\textsuperscript{4}\textbf{,} 
    \textbf{Haoyu Song}\textsuperscript{5}\textbf{,} 
    \textbf{Duyu Tang}\textsuperscript{5}\textbf{,} 
    \textbf{Hongyu Zhang}\textsuperscript{6}\textbf{,} 
    \textbf{Zibin Zheng}\textsuperscript{1}
    \\
    \textsuperscript{1}Sun Yat-sen University,
    \textsuperscript{2}Independent Researcher, 
    \textsuperscript{3}Hangzhou Normal University,\\
    \textsuperscript{4}Zhejiang University,
    \textsuperscript{5}Huawei Technologies Co, Ltd,
    \textsuperscript{6}Chongqing University
    \\
    \texttt{\{lich535, guolh8, guody5\}@mail2.sysu.edu.cn}, 
    \texttt{\{wangylin36, zhzibin\}@mail.sysu.edu.cn},\\
    \texttt{wtao@ieee.org},
    \texttt{20100119@hznu.edu.cn},
    \texttt{chenjiachi317@gmail.com}, \\
    \texttt{\{songhaoyu1, tangduyu\}@huawei.com},
    \texttt{hyzhang@cqu.edu.cn}
}

\begin{document}

\maketitle

\begin{abstract}
Issue resolution, a complex Software Engineering (SWE) task integral to real-world development, has emerged as a compelling challenge for artificial intelligence. The establishment of benchmarks like SWE-bench revealed this task as profoundly difficult for large language models, thereby significantly accelerating the evolution of autonomous coding agents. This paper presents a systematic survey of this emerging domain. We begin by examining data construction pipelines, covering automated collection and synthesis approaches. We then provide a comprehensive analysis of methodologies, spanning training-free frameworks with their modular components to training-based techniques, including supervised fine-tuning and reinforcement learning. Subsequently, we discuss critical analyses of data quality and agent behavior, alongside practical applications. Finally, we identify key challenges and outline promising directions for future research. An open-source repository is maintained at \url{https://github.com/DeepSoftwareAnalytics/Awesome-Issue-Resolution} to serve as a dynamic resource in this field.
\end{abstract}

\section{Introduction}
\begin{figure}[t]  
  \includegraphics[width=\columnwidth]{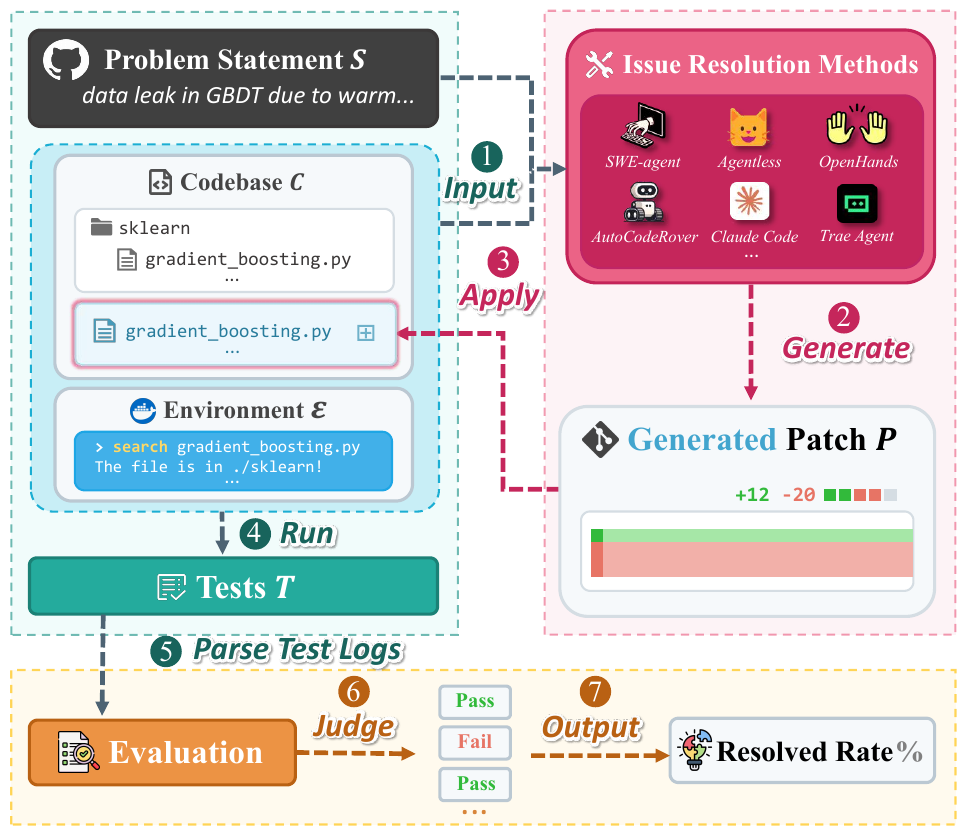}  
  \caption{Issue resolution task.}  
  \label{fig:task}
\end{figure}

The vision of constructing a true AI software engineer has long been an appealing prospect in computer science. In pursuit of this, researchers initially relied on function-level code generation benchmarks such as HumanEval~\cite{chen2021evaluatinglargelanguagemodels}. Driven by the remarkable success of Large Language Models (LLMs), the prospect of automating software engineering to function akin to a human developer has become increasingly attainable. However, they often struggle to handle the dynamic interactions with development environments and human collaboration in realistic scenarios. To address this limitation and align evaluation with authentic software development scenarios, \citet{jimenez2024swebenchlanguagemodelsresolve} introduced SWE-bench and defined the task of issue resolution. This task requires an automatic approach to help LLMs navigate complex, multi-file repositories to resolve issues (see Figure~\ref{fig:task} and Section~\ref{sec:background}). 
By revealing the difficulty of repository-level engineering, SWE-bench catalyzed a research frontier focused on navigating and modifying environments~\cite{pan2025trainingsoftwareengineeringagents}. It marks a departure from initial software generation as explored in ChatDev~\cite{qian2024chatdevcommunicativeagentssoftware} and MetaGPT~\cite{hong2024metagptmetaprogrammingmultiagent}, to the subsequent stages of software maintenance and evolution.

Despite the surge of research in this new frontier, the literature remains fragmented. Current surveys primarily focus on code generation, failing to address the far more complex challenge of issue resolution. This paper aims to bridge this gap by providing the first in-depth survey of this domain.

We conducted a comprehensive survey of publicly available literature, including 175 papers and online resources relevant to issue resolution. We established a tailored classification framework to provide a structured perspective on this rapidly evolving domain. Consequently, our contributions can be summarized as follows: We present a \textbf{systematic survey} on issue resolution, organized by a \textbf{structured taxonomy} covering Data, Methods, and Analysis. Furthermore, we identify \textbf{key challenges and future directions}, and provide a \textbf{open-source repository} to support the community.

\tikzstyle{my-box}=[
    rectangle,
    draw=hidden-draw,
    rounded corners,
    text opacity=1,
    minimum height=2em,
    minimum width=5em,
    inner sep=2pt,
    align=center,
    line width=0.8pt,
]
\tikzstyle{leaf}=[my-box, minimum height=2em,
    fill=hidden-pink!80, text=black, align=left,font=\huge,
    inner xsep=8pt,
    inner ysep=6pt,
    line width=0.8pt,
]
\tikzstyle{group-leaf}=[leaf, fill=hidden-pink!60]
\tikzstyle{last-leaf}=[leaf, fill=hidden-pink!40]

\begin{figure*}[t!]
    \vspace{-1.0cm}
    \centering
    \resizebox{\textwidth}{!}{%
        \begin{forest}
            forked edges,
            for tree={
                grow=east,
                reversed=true,
                anchor=base west,
                parent anchor=east,
                child anchor=west,
                base=left,
                font=\Huge,
                rectangle,
                draw=hidden-draw,
                rounded corners,
                align=left,
                minimum width=4em,
                edge+={darkgray, line width=1pt},
                s sep=4pt,
                inner xsep=8pt,
                inner ysep=6pt,
                line width=0.8pt,
                ver/.style={rotate=90, child anchor=north, parent anchor=south, anchor=center},
            },
            where level=1{font=\Huge,}{},
            where level=2{font=\Huge,}{},
            where level=3{font=\Huge,}{},
            where level=4{font=\Huge,}{},
            [
                Issue Resolution, ver
                [
                    Data\\(\S \ref{sec:data})
                    [
                        Datasets\\(\S \ref{sec:datasets})
                        [
                            {Evaluation Datasets\\(\S \ref{sec:evaluation-datasets}), Table~\ref{tab:dataset-survey}}
                            [SWE-bench~\cite{jimenez2024swebenchlanguagemodelsresolve}{, }
                            SWE-bench Lite~\cite{jimenez2024swebenchlanguagemodelsresolve}{, }
                            SWE-bench Verified~\cite{OpenAI2024swev}{, }\\
                             SWE-bench-java~\cite{zan2024swebenchjavagithubissueresolving}{, }
                             Visual SWE-bench~\cite{zhang-etal-2025-codev}{, }
                             SWE-Lancer~\cite{miserendino2025swelancerfrontierllmsearn}{, }\\
                             FEA-Bench~\cite{li2025feabenchbenchmarkevaluatingrepositorylevel}{, }
                             Multi-SWE-bench~\cite{zan2025multiswebenchmultilingualbenchmarkissue}{, }
                             SWE-PolyBench~\cite{rashid2025swepolybenchmultilanguagebenchmarkrepository}{, }\\
                             SWE-bench Multilingual~\cite{yang2025swesmithscalingdatasoftware}{, }
                             SwingArena~\cite{xu2025swingarenacompetitiveprogrammingarena}{, }
                             SWE-bench Multimodal~\cite{yang2024swebenchmultimodalaisystems}{, }\\
                             OmniGIRL~\cite{guo2025omnigirl}{, }
                             SWE-bench-Live~\cite{zhang2025swebenchgoeslive}{, }
                             SWE-Factory~\cite{guo2025swefactoryautomatedfactoryissue}{, }\\
                             SWE-MERA~\cite{adamenko2025swemeradynamicbenchmarkagenticly}{, }
                             SWE-Perf~\cite{he2025sweperflanguagemodelsoptimize}{, }
                             SWE-Bench Pro~\cite{deng2025swebenchproaiagents}{, }\\
                             SWE-InfraBench~\cite{tarasova2025sweinfrabench}{, }
                             SWE-Sharp-Bench~\cite{mhatre2025swesharpbenchreproduciblebenchmarkc}{, }
                             SWE-fficiency~\cite{ma2025swefficiencylanguagemodelsoptimize}{, }\\
                             SWE-Compass~\cite{xu2025swecompassunifiedevaluationagentic}{, }
                             SWE-Bench++~\cite{wang2025swebenchframeworkscalablegeneration}{, }
                             SWE-EVO~\cite{thai2025sweevobenchmarkingcodingagents}.
                             , group-leaf]
                        ]
                        [
                            Training Datasets\\(\S \ref{sec:training-datasets})
                            [SWE-bench-train~\cite{jimenez2024swebenchlanguagemodelsresolve}{, }
                            SWE-bench-extra~\cite{jimenez2024swebenchlanguagemodelsresolve}{, }
                            Multi-SWE-RL~\cite{zan2025multiswebenchmultilingualbenchmarkissue}{, }\\
                            R2E-Gym~\cite{jain2025r2egymproceduralenvironmentshybrid}{, }
                            SWE-Synth~\cite{pham2025swesynthsynthesizingverifiablebugfix}{, }
                            LocAgent~\cite{yu2025orcalocallmagentframework}{, }
                            SWE-Smith~\cite{yang2025swesmithscalingdatasoftware}{, }\\
                            SWE-Fixer~\cite{xie2025swefixertrainingopensourcellms}{, }
                            SWELoc~\cite{reddy2025sweranksoftwareissuelocalization}{, }
                            SWE-Gym~\cite{pan2025trainingsoftwareengineeringagents}{, }
                            SWE-Flow~\cite{zhang2025sweflowsynthesizingsoftwareengineering}{, }\\
                            SWE-Factory~\cite{guo2025swefactoryautomatedfactoryissue}{, }
                            Skywork-SWE~\cite{zeng2025skyworksweunveilingdatascaling}{, }
                            RepoForge~\cite{chen2025repoforgetrainingsotafastthinking}{, }
                            SWE-Mirror~\cite{wang2025swemirrorscalingissueresolvingdatasets}{, }\\
                            SWE-Bench++~\cite{wang2025swebenchframeworkscalablegeneration}{, }
                            SWE-Lego~\cite{tao2026swelegopushinglimitssupervised}.
                            , group-leaf]
                        ]
                    ]
                    [
                        Data Construction\\(\S \ref{sec:data-construction})
                        [
                            Automated data\\Collection (\S \ref{sec:data-collection})
                            [SWE-rebench~\cite{badertdinov2025swerebenchautomatedpipelinetask}{, }
                            RepoLaunch~\cite{zhang2025swebenchgoeslive}{, }
                            SWE-Factory~\cite{guo2025swefactoryautomatedfactoryissue}{, }\\
                            SWE-MERA~\cite{adamenko2025swemeradynamicbenchmarkagenticly}{, }
                            RepoForge~\cite{chen2025repoforgetrainingsotafastthinking}{, }
                            Multi-Docker-Eval~\cite{fu2025multidockerevalshovelgoldrush}.
                            , group-leaf]
                        ]
                        [
                            Automated data\\Synthesis (\S \ref{sec:data-synthesis})
                            [
                            Learn-by-interact~\cite{su2025learnbyinteract}{, }
                            R2E-Gym~\cite{jain2025r2egymproceduralenvironmentshybrid}{, }
                            SWE-Synth~\cite{pham2025swesynthsynthesizingverifiablebugfix}{, }\\
                            SWE-smith~\cite{yang2025swesmithscalingdatasoftware}{, }
                            SWE-Flow~\cite{zhang2025sweflowsynthesizingsoftwareengineering}{, }
                            SWE-Mirror~\cite{wang2025swemirrorscalingissueresolvingdatasets}.
                            , group-leaf]
                        ]
                    ]
                ]
                [
                    Methods\\(\S \ref{sec:methods})
                    [
                        Training-free\\Methods\\(\S \ref{sec:training-free-methods})
                            [
                                Frameworks\\(\S \ref{sec:frameworks})
                                    [
                                        Single-agent\\(\S \ref{sec:single-agent})
                                        [SWE-agent~\cite{yang2024sweagentagentcomputerinterfacesenable}{, }
                                        Aider~\cite{aider2026aider}{, }
                                        Devin~\cite{cognitionteam2025devin}{, }
                                        PatchPilot~\cite{li2025patchpilotcostefficientsoftwareengineering}{, }\\
                                        LCLM~\cite{jiang2025puttingcontextsimplifyingagents}{, }
                                        DGM~\cite{zhang2025darwingodelmachineopenended}{, }
                                        Trae Agent~\cite{traeresearchteam2025traeagentllmbasedagent}{, }\\
                                        SE-Agent~\cite{lin2025seagentselfevolutiontrajectoryoptimization}{, }
                                        Lita~\cite{dai2025litalightagentuncovers}{, }
                                        TOM-SWE~\cite{zhou2025tomsweusermentalmodeling}{, }
                                        Live-SWE-agent~\cite{lin2025seagentselfevolutiontrajectoryoptimization}{, }\\
                                        Confucius Code Agent~\cite{wong2025confuciuscodeagentscalable}.
                                        , group-leaf]
                                    ]
                                    [   
                                        Multi-agent\\(\S \ref{sec:multi-agent})
                                        [MAGIS~\cite{tao2024magisllmbasedmultiagentframework}{, }
                                        AutoCodeRover~\cite{zhang2024autocoderoverautonomousprogramimprovement}{, }
                                        CodeR~\cite{chen2024coderissueresolvingmultiagent}{, }
                                        OpenHands~\cite{wang2025openhandsopenplatformai}{, }\\
                                        AgentScope~\cite{doc2026swe}{, }
                                        OrcaLora~\cite{yu2025orcalocallmagentframework}{, }
                                        DEI~\cite{zhang2024diversityempowersintelligenceintegrating}{, }
                                        MarsCode Agent~\cite{liu2024marscodeagentainativeautomated}{, }\\
                                        Lingxi~\cite{github2026lingxi}{, }
                                        Devlo~\cite{devlo2026devlo}{, }
                                        Refact.ai Agent~\cite{refact2026ai}{, }
                                        HyperAgent~\cite{phan2025hyperagentgeneralistsoftwareengineering}{, }\\
                                        SWE-Search~\cite{antoniades2025swesearchenhancingsoftwareagents}{, }
                                        CodeCoR~\cite{pan2025codecorllmbasedselfreflectivemultiagent}{, }
                                        Agent KB~\cite{tang2025agentkbleveragingcrossdomain}{, }
                                        SWE-Debate~\cite{li2025swedebatecompetitivemultiagentdebate}{, }\\
                                        SWE-Exp~\cite{chen2025sweexpexperiencedrivensoftwareissue}{, }
                                        Meta-RAG~\cite{tawosi2025metaraglargecodebasesusing}{, }
                                        , group-leaf]
                                    ]
                                    [
                                        Workflow\\(\S \ref{sec:workflow})
                                        [
                                            Agentless~\cite{xia2024agentlessdemystifyingllmbasedsoftware}{, }
                                            Conversational Pipeline~\cite{cheshkov2024exploringpotentialconversationaltest}{, }
                                            SynFix~\cite{tang-etal-2025-synfix}{, }\\
                                            CodeV~\cite{zhang-etal-2025-codev}{, }
                                            GUIRepair~\cite{huang2025seeingfixingcrossmodalreasoning}.
                                        , group-leaf]
                                    ]
                            ]
                            [
                                Modules\\(\S \ref{sec:modules})
                                    [
                                        Tool\\(\S \ref{sec:tool})
                                        [
                                        MAGIS~\cite{tao2024magisllmbasedmultiagentframework}{, }
                                        AutoCodeRover~\cite{zhang2024autocoderoverautonomousprogramimprovement}{, }
                                        SWE-agent~\cite{yang2024sweagentagentcomputerinterfacesenable}{, }
                                        Alibaba LingmaAgent~\cite{ma2025alibabalingmaagentimprovingautomated}{, }\\
                                        OpenHands~\cite{wang2025openhandsopenplatformai}{, }
                                        SpecRover~\cite{ruan2024specrovercodeintentextraction}{, }
                                        MarsCode Agent~\cite{liu2024marscodeagentainativeautomated}{, }
                                        RepoGraph~\cite{ouyang2025repographenhancingaisoftware}{, }\\
                                        SuperCoder2.0~\cite{gautam2024supercoder20technicalreportexploring}{, }
                                        EvoCoder~\cite{lin2024llmscontinuouslearnersimproving}{, }
                                        AEGIS~\cite{wang2025aegisagentbasedframeworkgeneral}{, }
                                        CoRNStack~\cite{suresh2025cornstack}{, }\\
                                        OrcaLoca~\cite{yu2025orcalocallmagentframework}{, }
                                        DARS~\cite{aggarwal2025darsdynamicactionresampling}{, }
                                        Otter~\cite{ahmed2025ottergeneratingtestsissues}{, }
                                        Quadropic Insiders~\cite{insiders2026quadropic}{, }\\
                                        Issue2Test~\cite{nashid2025issue2testgeneratingreproducingtest}{, }
                                        KGCompass~\cite{yang2025enhancingrepositorylevelsoftwarerepair}{, }
                                        CoSIL~\cite{jiang2025issuelocalizationllmdriveniterative}{, }
                                        InfantAgent-Next~\cite{lei2025infantagentnextmultimodalgeneralistagent}{, }\\
                                        Co-PatcheR~\cite{tang2025copatchercollaborativesoftwarepatching}{, }
                                        SWERank~\cite{reddy2025sweranksoftwareissuelocalization}{, }
                                        Nemotron-CORTEXA~\cite{sohrabizadeh2025nemotroncortexa}{, }\\
                                        LCLM~\cite{jiang2025puttingcontextsimplifyingagents}{, }
                                        SACL~\cite{gupta2025saclunderstandingcombatingtextual}{, }
                                        SWE-Debate~\cite{li2025swedebatecompetitivemultiagentdebate}{, }
                                        OpenHands-Versa~\cite{soni2025codingagentsmultimodalbrowsing}{, }\\
                                        SemAgent~\cite{pabba2025semagentsemanticsawareprogram}{, }
                                        Repeton~\cite{vinh2025repetonstructuredbugrepair}{, }
                                        cAST~\cite{zhang2025castenhancingcoderetrievalaugmented}{, }
                                        Prometheus~\cite{chen2025prometheusunifiedknowledgegraphs}{, }\\
                                        Git Context Controller~\cite{wu2025gitcontextcontrollermanage}{, }
                                        Trae Agent~\cite{traeresearchteam2025traeagentllmbasedagent}{, }
                                        BugPilot~\cite{sonwane2025bugpilotcomplexbuggeneration}{, }
                                        TestPrune~\cite{chen2025oldmeetsnewevaluating}{, }\\
                                        e-Otter++~\cite{ahmed2025executionfeedbackdriventestgeneration}{, }
                                        Meta-RAG~\cite{tawosi2025metaraglargecodebasesusing}{, }
                                        InfCode~\cite{li2025infcodeadversarialiterativerefinement}{, }
                                        GraphLocator~\cite{liu2025graphlocatorgraphguidedcausalreasoning}.
                                        , group-leaf]
                                    ]
                                    [
                                        Memory\\(\S \ref{sec:memory})
                                        [
                                        Infant Agent~\cite{lei2024infantagenttoolintegratedlogicdriven}{, }
                                        EvoCoder~\cite{lin2024llmscontinuouslearnersimproving}{, }
                                        Learn-by-interact~\cite{su2025learnbyinteract}{, }
                                        DGM~\cite{zhang2025darwingodelmachineopenended}{, }\\
                                        ExpeRepair~\cite{mu2025experepairdualmemoryenhancedllmbased}{, }
                                        Agent KB~\cite{tang2025agentkbleveragingcrossdomain}{, }
                                        SWE-Exp~\cite{chen2025sweexpexperiencedrivensoftwareissue}{, }
                                        RepoMem~\cite{wang2025improvingcodelocalizationrepository}{, }\\
                                        AgentDiet~\cite{xiao2025improvingefficiencyllmagent}{, }
                                        ReasoningBank~\cite{ouyang2025reasoningbankscalingagentselfevolving}{, }
                                        MemGovern~\cite{wang2026memgovernenhancingcodeagents}.
                                        , group-leaf]
                                    ]
                            ]
                            [
                                Inference-time\\Scaling (\S \ref{sec:inference-time-scaling})
                                [
                                SWE-Search~\cite{antoniades2025swesearchenhancingsoftwareagents}{, }
                                CodeMonkeys~\cite{ehrlich2025codemonkeysscalingtesttimecompute}{, }
                                SWE-PRM~\cite{gandhi2025agentsastraycoursecorrectingswe}{, }\\
                                ReasoningBank~\cite{ehrlich2025codemonkeysscalingtesttimecompute}{, }
                                SIADAFIX~\cite{cao2025siadafixissuedescriptionresponse}.
                                , group-leaf]
                            ]
                    ]
                    [
                        Training-based\\Methods\\(\S \ref{sec:training-methods})
                        [SFT-based\\Methods\\(\S \ref{sec:sft-methods})
                            [
                            Lingma SWE-GPT~\cite{ma2024lingmaswegptopendevelopmentprocesscentric}{, }
                            ReSAT~\cite{ma2024repositorystructureawaretrainingmakes}{, }
                            Scaling data collection~\cite{nebius2024scaling}{, }
                            CodeXEmbed~\cite{liu2025codexembed}{, }\\
                            SWE-Gym~\cite{pan2025trainingsoftwareengineeringagents}{, }
                            Thinking Longer~\cite{ma2025thinkinglongerlargerenhancing}{, }
                            Search for training~\cite{zainullina2025guidedsearchstrategiesnonserializable}{, }
                            Co-PatcheR~\cite{tang2025copatchercollaborativesoftwarepatching}{, }\\
                            MCTS-Refined CoT~\cite{wang2025mctsrefinedcothighqualityfinetuning}{, }
                            SWE-Swiss~\cite{He2025SWESwiss}{, }
                            Devstral~\cite{rastogi2025devstralfinetuninglanguagemodels}{, }
                            Kimi-Dev~\cite{yang2025kimidevagentlesstrainingskill}{, }\\
                            SWE-Compressor~\cite{liu2025contexttoolcontextmanagement}{, }
                            SWE-Lego~\cite{tao2026swelegopushinglimitssupervised}{, }
                            SWE-rebench(SFT)~\cite{trofimova2025openhandstrajs}{, }
                            Agentic Rubrics~\cite{raghavendra2026agenticrubricscontextualverifiers}.
                            , group-leaf]
                        ]
                        [RL-based\\Methods\\(\S \ref{sec:rl-methods})
                            [
                            SWE-RL~\cite{wei2025swerladvancingllmreasoning}{, }
                            SoRFT~\cite{ma2025sorftissueresolvingsubtaskoriented}{, }
                            SEAlign~\cite{zhang2025sealignalignmenttrainingsoftware}{, }
                            SWE-Dev$_1$~\cite{du2025swedevevaluatingtrainingautonomous}{, }\\
                            Satori-SWE~\cite{zeng2025satorisweevolutionarytesttimescaling}{, }
                            Agent-RLVR~\cite{da2025agentrlvrtrainingsoftwareengineering}{, }
                            DeepSWE~\cite{luo2025deepswe}{, }
                            SWE-Dev$_2$~\cite{wang2025swedevbuildingsoftwareengineering}{, }\\
                            Tool-integrated RL~\cite{ma2025toolintegratedreinforcementlearningrepo}{, }
                            SWE-Swiss~\cite{He2025SWESwiss}{, }
                            SeamlessFlow~\cite{wang2025seamlessflowtraineragentisolation}{, }
                            DAPO~\cite{golubev2025traininglongcontextmultiturnsoftware}{, }\\
                            CoreThink~\cite{vaghasiya2025corethinksymbolicreasoninglayer}{, }
                            CWM~\cite{faircodegenteam2025cwmopenweightsllmresearch}{, }
                            EntroPO~\cite{yu2025buildingcodingagentsentropyenhanced}{, }
                            Kimi-Dev~\cite{yang2025kimidevagentlesstrainingskill}{, }\\
                            FoldGRPO~\cite{sun2025scalinglonghorizonllmagent}{, }
                            GRPO-based Method~\cite{wang2025practitionersguidemultiturnagentic}{, }
                            TSP~\cite{xiong2025think}{, }
                            Self-play SWE-RL~\cite{wei2025trainingsuperintelligentsoftwareagents}{, }\\
                            SWE-Playground~\cite{zhu2026trainingversatilecodingagents}{, }
                            Supervised RL~\cite{deng2025supervisedreinforcementlearningexpert}{, }
                            OSCA~\cite{zhang-etal-2025-scaling}{, }
                            SWE-RM~\cite{shum2025swermexecutionfreefeedbacksoftware}{, }\\
                            One Tool Is Enough~\cite{zhang2026toolenoughreinforcementlearning}{, }
                            Let It Flow~\cite{wang2026letflowagenticcrafting}{, }
                            KAT-Coder~\cite{zhan2025katcodertechnicalreport}{, }
                            Seed1.5-Thinking ~\cite{seed2025seed15thinkingadvancingsuperbreasoning}{, }\\
                            Deepseek V3.2~\cite{deepseekai2025deepseekv32pushingfrontieropen}{, }
                            Kimi-K2-Instruct~\cite{kimiteam2025kimik2openagentic}{, }
                            gpt-oss-120b \& gpt-oss-20b~\cite{agarwal2025gpt}{, }\\
                            Qwen3-Coder~\cite{yang2025qwen3technicalreport}{, }
                            GLM-4.6~\cite{zeng2025glm,agarwal2025gpt}{, }
                            Minimax M2~\cite{chen2025minimax}{, }\\
                            LongCat-Flash-Think~\cite{meituanlongcatteam2025introducinglongcatflashthinkingtechnicalreport}{, }
                            MiMo-V2-Flash~\cite{coreteam2026mimov2flashtechnicalreport}.
                            , group-leaf]
                        ]
                    ]
                ]
                [
                    Analysis\\(\S \ref{sec:analysis})
                    [
                        Data Analysis\\(\S \ref{sec:data-quality})
                        [
                        SWE-bench Verified~\cite{OpenAI2024swev}{, }
                        SWE-Bench+~\cite{aleithan2024swebenchenhancedcodingbenchmark}{, }
                        Patch Correctness~\cite{wang_are_2025}{, }
                        UTBoost~\cite{yu2025utboostrigorousevaluationcoding}{, }\\
                        Trustworthiness~\cite{mathews2025automatedsoftwareengineertrustworthy}{, }
                        Rigorous agentic benchmarks~\cite{zhu2025establishingbestpracticesbuilding}{, }
                        The SWE-Bench Illusion~\cite{liang2025swebenchillusionstateoftheartllms}{, }\\
                        Revisiting SWE-Bench~\cite{Aleithan2025revisitingswebench}{, }
                        SPICE~\cite{oliva2025spiceautomatedswebenchlabeling}{, }
                        Data contamination~\cite{prathifkumar2025doesswebenchverifiedtestagent}.
                        , group-leaf]
                    ]
                    [
                        Methods Analysis\\(\S \ref{sec:method-analysis})
                        [
                        Context Retrieval~\cite{kovrigin2024importancereasoningcontextretrieval}{, }
                        Evaluating software development agents~\cite{chen2024evaluatingsoftwaredevelopmentagents}{, }
                        Overthinking~\cite{cuadron2025dangeroverthinkingexaminingreasoningaction}{, }\\
                        Beyond final code~\cite{chen2025finalcodeprocessorientederror}{, }
                        GSO~\cite{shetty2025gsochallengingsoftwareoptimization}{, }
                        Dissecting the SWE-Bench Leaderboards~\cite{martinez2025dissectingswebenchleaderboardsprofiling}{, }\\
                        Security analysis~\cite{sajadi2025aigeneratedfixessecureanalyzing}{, }
                        Failures analysis~\cite{liu2025empiricalstudyfailuresautomated}{, }
                        SeaView~\cite{bula2025seaviewsoftwareengineeringagent}{, }
                        SWEnergy~\cite{tripathy2026swenergy}{, }\\
                        Strong-Weak Model Collaboration~\cite{gandhi2025empiricalstudystrongweakmodel}{, }
                        Agents in the Wild~\cite{logicstarai2025agentsinthewild}.
                        , group-leaf]
                    ]
                ]
            ]
        \end{forest}
    }
    \caption{Overall perspective on data, methods, and analysis for SWE tasks, featuring corresponding papers. }
    \label{fig:outline_mindmap}
\end{figure*}
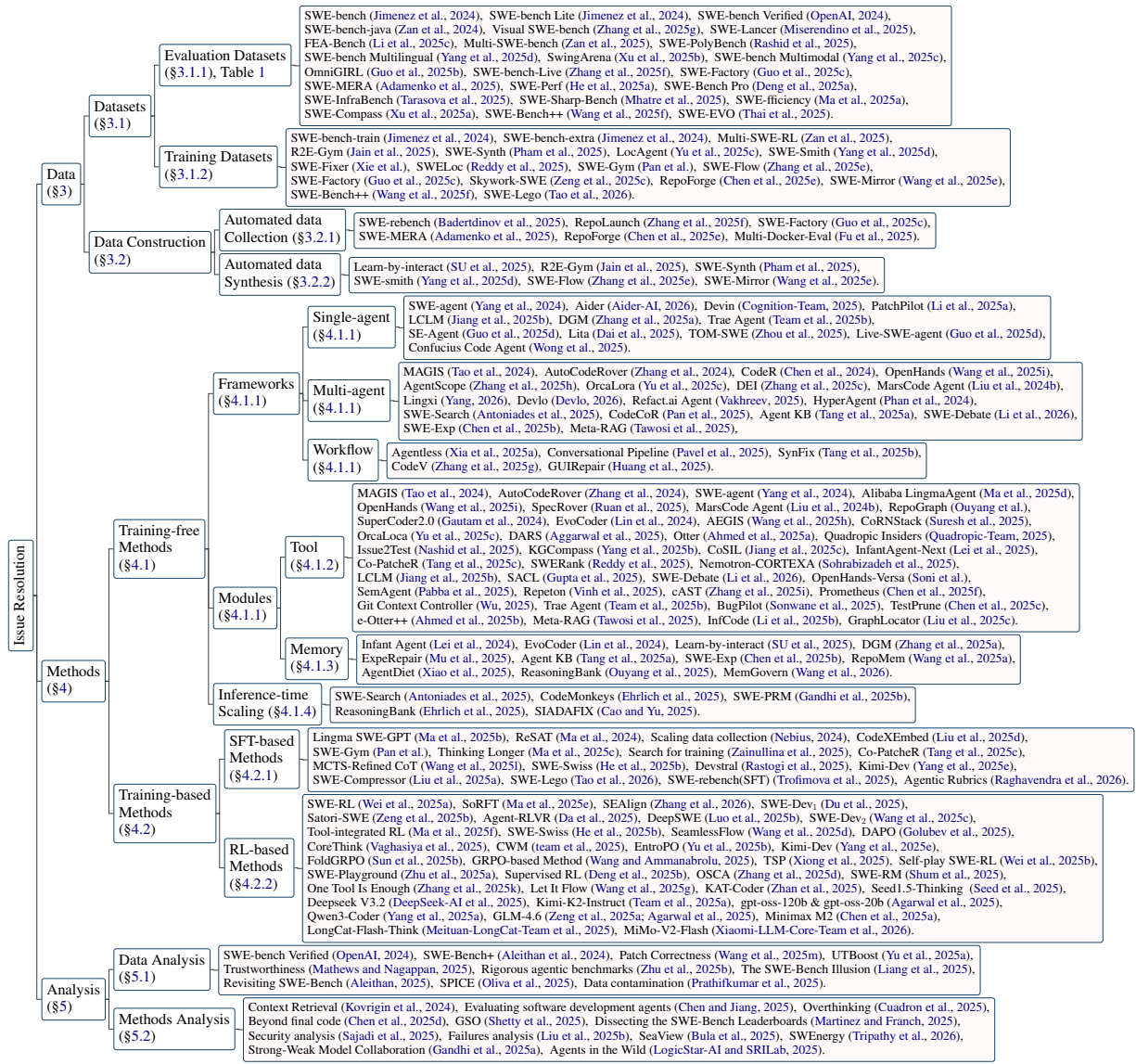

\section{Task formulation}
\label{sec:background}

Issue resolution requires LLMs to synthesize a valid code change (also called a patch) $\gP$ to resolve the issue, interacting with a codebase (as shown in Figure~\ref{fig:task}). 
Formally, an instance of the task can be expressed as $\gI = (\gD, \gC, \gT)$, comprising an issue description $\gD$, the codebase $\gC$, and corresponding tests $\gT$. The observable parts of the instance, $\gD$ and $\gC$, are available during the resolving process, which contains supplementary information on the corresponding environment $\gE$ that can be explored. So a method $\mathcal{M}$ is expected to achieve:

\begin{equation}
    \gP = \gM(\gD,\gC,\gE)
\end{equation}

After the patch $\gP$ is applied to $\gC$, the evaluation is conducted through running tests on the modified codebase $\gC'=\text{Apply}(\gC, \gP)$. The resolution outcome $r\in\{0, 1\}$ is then determined by the execution of $\gT$, denoted as $\text{Exec}(\gC', \gT)$. On a dataset including $n$ instances($\sI = \{ \gI_i \}_{i=1}^{N}$), overall performance metric is:

\begin{equation}
    \text{Resolved Rate} = \frac{1}{|\sI|} \sum_{i=1}^{|\sI|} r_i
\end{equation}

\begin{figure*}[t]
  \centering    
  \includegraphics[width=1\textwidth]{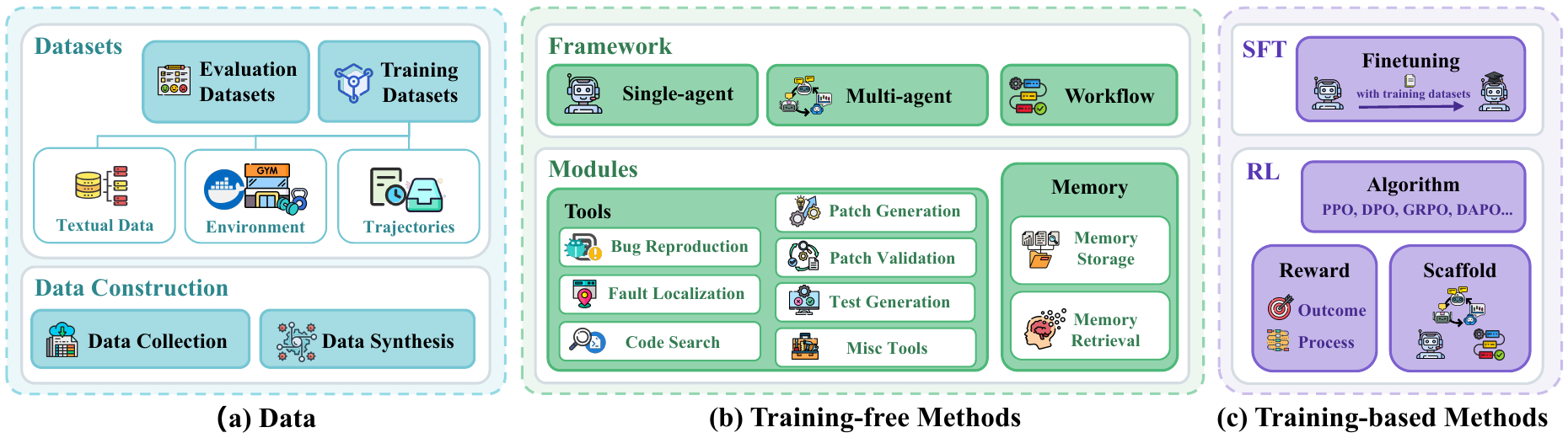}
  \caption{A classification overview of data, training-free methods, and training methods for solving SWE tasks. }
  \label{fig:framework}
\end{figure*}

\section{Data}
\label{sec:data}
Data is fundamental to the issue-resolution task, serving as both an evaluation benchmark and a training resource. Thus, datasets are classified into evaluation (\S~\ref{sec:evaluation-datasets}) and training (\S~\ref{sec:training-datasets}) sets. Construction approaches are divided into two types: data collection (\S~\ref{sec:data-collection}) from real-world online sources, and data synthesis (\S~\ref{sec:data-synthesis}) achieved by rewriting real-world data or rule-based generation. Statistics for all datasets are provided in \S~\ref{sec:data-detail}.

\subsection{Datasets}
\label{sec:datasets}

\subsubsection{Evaluation datasets}
\label{sec:evaluation-datasets}
SWE-bench~\cite{jimenez2024swebenchlanguagemodelsresolve} established the datasets by collecting issue–Pull Request (PR) pairs from popular Python repositories, pairing each issue with a full repository snapshot for resolution. However, invalid tests and underspecified descriptions in the original dataset make many instances unsuitable for evaluation. To ensure data quality, SWE-bench Verified~\cite{OpenAI2024swev} was introduced, offering a subset of manually validated samples as a trusted benchmark.

While most evaluation datasets target Python, researchers extend these tasks to ten different programming languages to broaden the linguistic scope, as seen in SWE-bench Multilingual and Multi-SWE-bench, et al.~\cite{zan2024swebenchjavagithubissueresolving,zan2025multiswebenchmultilingualbenchmarkissue,rashid2025swepolybenchmultilanguagebenchmarkrepository,guo2025omnigirl,yang2025swesmithscalingdatasoftware,mhatre2025swesharpbenchreproduciblebenchmarkc}. 
To address the limitation of relying solely on textual data, researchers have focused on aggregating multimodal information, primarily derived from images such as UI screenshots and diagrams~\cite{yang2024swebenchmultimodalaisystems,zhang-etal-2025-codev,guo2025omnigirl}. For instance, \citet{yang2024swebenchmultimodalaisystems} integrates these visual contexts and introduces a novel visual tester to validate the correctness of visual modifications.

To bridge the gap between initial evaluation datasets and realistic software development scenarios, researchers introduced datasets~\cite{miserendino2025swelancerfrontierllmsearn,deng2025swebenchproaiagents,tarasova2025sweinfrabench} that incorporate enterprise-level complexity and diverse domains. Furthermore, the scope of issue resolution tasks has been expanded to broader vision on software engineering, such as incremental development~\cite{li2025feabenchbenchmarkevaluatingrepositorylevel} and long-horizon software evolution~\cite{zeng2025satorisweevolutionarytesttimescaling}. Recent works have also shifted towards refining metrics for issue resolution, such as efficiency and safety~\cite{he2025sweperflanguagemodelsoptimize,ma2025swefficiencylanguagemodelsoptimize,xu2025swecompassunifiedevaluationagentic}.

\subsubsection{Training datasets}
\label{sec:training-datasets}

\paragraph{Textual data} The initial training data typically consists of raw task instances in the form of static issue-PR, as exemplified by the training set in SWE-bench~\cite{jimenez2024swebenchlanguagemodelsresolve}, to equip models with fundamental capabilities to resolve issues before they engage with interactive environments.

\paragraph{Training Environment} Environment datasets aim to address the limitations of static text by constructing interactive code environments that provide LLMs with execution feedback. In early initiatives, researchers attempted to equip each task instance with a corresponding Conda or Docker environment, enabling LLMs to incorporate code execution results as feedback during the training process, as seen in benchmarks like Multi-SWE-RL~\cite{zan2025multiswebenchmultilingualbenchmarkissue}. Nevertheless, these datasets often overlook the specific interface design required for effective LLM-environment interaction. To address this, \citet{jain2025r2egymproceduralenvironmentshybrid} introduced a more interactive Gym environment (R2E-Gym) that utilizes LLM-synthesized test cases to verify environment usability, thereby synthesizing large-scale environment data. Similarly, \citet{pan2025trainingsoftwareengineeringagents} constructed Gym environments based on real-world GitHub issues.

\paragraph{Trajectories}
Trajectory datasets capture the procedural interplay between LLMs and execution environments through tool invocations and feedback loops~\cite{nebius2024scaling,jain2025r2egymproceduralenvironmentshybrid, pan2025trainingsoftwareengineeringagents,pham2025swesynthsynthesizingverifiablebugfix,xie2025swefixertrainingopensourcellms,guo2025swefactoryautomatedfactoryissue}(more details in \S~\ref{sec:data-detail}). To obtain high-quality trajectories, researchers typically employ inference-time scaling strategies to generate many candidates, then apply verifiers to filter and select the best~\cite{nebius2024scaling, yang2025swesmithscalingdatasoftware}.

\subsection{Data construction}
\label{sec:data-construction}

Training data construction for software agents is transitioning from manual processes to automated pipelines (see \S~\ref{sec:data-construction-appendix} for background details).

\subsubsection{Automated data collection}
\label{sec:data-collection}

Static datasets often suffer from rigidity, high maintenance costs, and limited scale, which impede the effective training of robust models. In response, the field is evolving towards scalable, automated data collection methods. Those automated pipelines usually leverage LLMs to explore repository configurations, identify files related to environment setup, and generate corresponding dependency installation commands to build Docker images for individual issues. Subsequently, they employ existing testing frameworks and predefined log parsers to analyze test execution results, encompassing both workflow-based and agent-based paradigms, such as SWE-rebench and RepoLaunch~\cite{badertdinov2025swerebenchautomatedpipelinetask,zhang2025swebenchgoeslive}. Notably, SWE-Factory employs a memory-enhanced framework for environment setup and verification, while utilizing an exit-code-based automatic grading method to design parsers capable of automatically interpreting execution results across diverse testing frameworks in different programming languages~\cite{guo2025swefactoryautomatedfactoryissue}. More recently, RepoForge~\cite{chen2025repoforgetrainingsotafastthinking} enhanced the pipeline's automation by incorporating a automatic verification mechanism based on SPICE~\cite{oliva2025spiceautomatedswebenchlabeling} following data construction, effectively replacing the need for human expert validation.

\subsubsection{Automated data synthesis}
\label{sec:data-synthesis}

To address limited real-world data and costly manual verification, researchers increasingly adopted automated data synthesis approaches. For instance, SWE-Synth~\cite{pham2025swesynthsynthesizingverifiablebugfix} rewrites target code and generates corresponding tests. Drawing on test-driven development, SWE-Flow~\cite{zhang2025sweflowsynthesizingsoftwareengineering} employs a runtime dependency graph to derive incremental code and requirements from unit tests. More recently, SWE-Smith~\cite{yang2025swesmithscalingdatasoftware} scales data by paraphrasing descriptions and injecting bugs, leveraging a shared environment to reduce storage overhead. Similarly, SWE-Mirror~\cite{wang2025swemirrorscalingissueresolvingdatasets} transplants real-world issues into target repositories to generate verifiable tasks under shared environments.

\section{Methods}
\label{sec:methods}

\subsection{Training-free method}
\label{sec:training-free-methods}
To overcome constraints such as fixed context windows and static knowledge, training-free methods utilizes external components and sophisticated prompting. As shown in Figure \ref{fig:framework}, we classify these methods into three categories based on the underlying framework: (1) \textit{Frameworks}, encompassing high-level architectures like single-agent, multi-agent, and fixed-workflow designs; (2) \textit{Modules}, providing plug-and-play augmentations such as \textit{Tools} for repository interaction and \textit{Memory} for experience accumulation; and (3) \textit{Inference-time Scaling} (or test-time scaling), employing search or parallelization strategies to enhance success rates without modifying model parameters.

\subsubsection{Frameworks}
\label{sec:frameworks}

To handle the multi-stage execution required for issue resolution, current research structures LLM activities into either dynamic agent-based or rigid workflow-based frameworks. 

\paragraph{Single-agent}
\label{sec:single-agent}

Analogous to software engineers who write code and invoke diverse tools to resolve issues, single-agent frameworks were initially constructed to execute tasks via tool-based interaction paradigms. SWE-agent~\cite{yang2024sweagentagentcomputerinterfacesenable} pioneers the agent-computer interface, which enables autonomous file navigation, code editing, and test execution, bridging natural language understanding with repository-level operations.

However, granting full autonomy for every decision often leads to redundant action sequences due to reasoning imprecision, resulting in prohibitive operational costs. To address it, \citet{li2025patchpilotcostefficientsoftwareengineering} reduces overhead by either constraining specific phases into rigid processes.

To further enhance generalization across diverse issue types, self-evolutionary frameworks have emerged to autonomously refine agent capabilities. For instance, Darwin Gödel Machine employs an evolutionary process starting from a minimal baseline, where the LLM generates, scores, and selects optimal candidate agent implementations over successive iterations to evolve its structure~\cite{zhang2025darwingodelmachineopenended,xia2025livesweagentsoftwareengineeringagents, lin2025seagentselfevolutiontrajectoryoptimization}.

\paragraph{Multi-agent}
\label{sec:multi-agent}

Introduced concurrently with single-agent systems, multi-agent frameworks focus on collaboration and task allocation, often performed in the form of human software development team~\cite{tao2024magisllmbasedmultiagentframework,zhang2024autocoderoverautonomousprogramimprovement,liu2024marscodeagentainativeautomated,antoniades2025swesearchenhancingsoftwareagents,yu2025orcalocallmagentframework}. For instance, MAGIS~\cite{tao2024magisllmbasedmultiagentframework} firstly implements this by assigning four agents customized for software evolution, enabling role-playing and autonomous meetings for effective communication.

However, those works largely relies on text-based contexts for information exchange and lacks explicit modeling of agent collaboration. To address this, CodeR~\cite{chen2024coderissueresolvingmultiagent} introduces task graphs that convert a high-level plan into a parsable, directed graph to ensure precise execution. Similarly, SWE-Debate~\cite{li2025swedebatecompetitivemultiagentdebate} adopts graph-based structures to orchestrate a three-round debate among specialized agents along code dependency traces, yielding more concrete solutions.

With the proliferation of diverse agent frameworks, recent research has shifted towards unified platforms capable of orchestrating collaboration among heterogeneous agents~\cite{wang2025openhandsopenplatformai,zhang2024diversityempowersintelligenceintegrating}. For example,~\citet{zhang2024diversityempowersintelligenceintegrating}. proposed DEIBase, which leverages LLMs to score and rank solutions generated by multiple agents, achieving superior performance over single-agent approaches.

\paragraph{Workflow}
\label{sec:workflow}
Workflow architectures improve stability by enforcing predefined steps instead of open-ended exploration. \citet{xia2024agentlessdemystifyingllmbasedsoftware} adopted a linear pipeline (localization, repair, and validation) to ensure efficiency and reproducibility. For visual tasks, researchers use vision language models to convert UI screenshots into code~\cite{huang2025seeingfixingcrossmodalreasoning} or textual descriptions~\cite{zhang-etal-2025-codev}. To handle complex codebases, \citet{tang-etal-2025-synfix} utilized dependency graphs to guide precise, repository-wide modifications instead of random search.

\label{sec:modules}
\subsubsection{Tool modules}
\label{sec:tool}

\begin{figure*}[t]
  \centering    
  \includegraphics[width=1\textwidth]{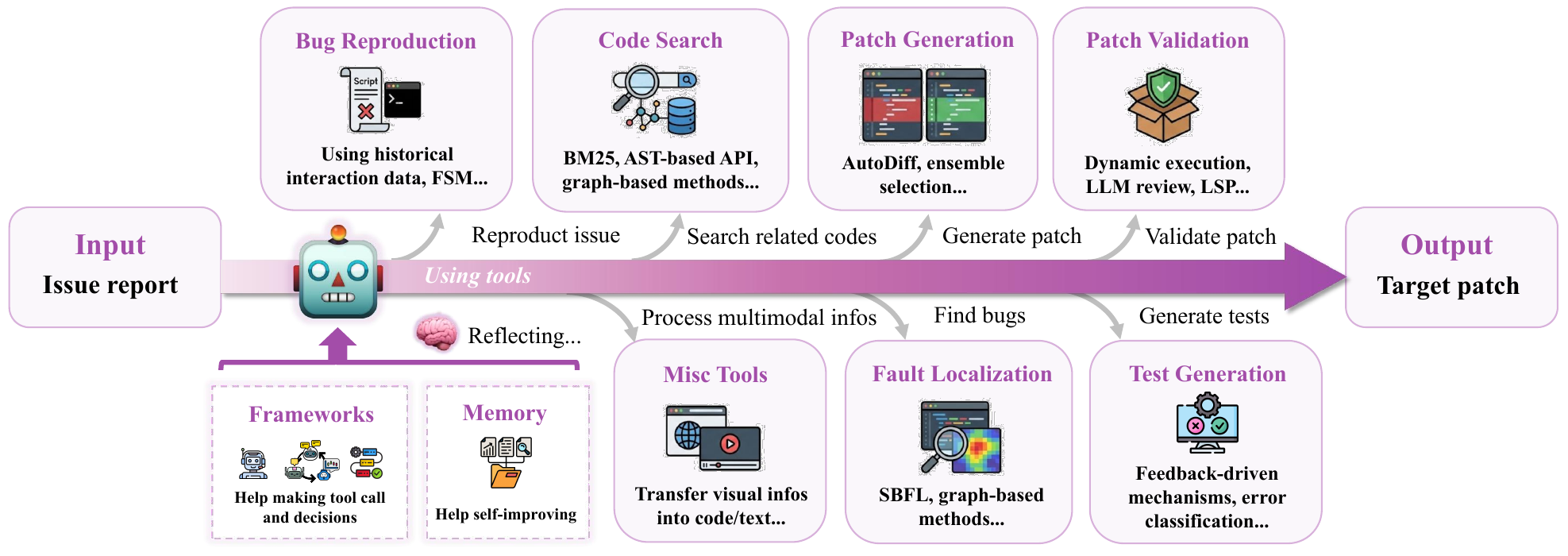}
  \caption{Taxonomy of tool modules for LLM-based issue resolution.}
  \label{fig:tool}
\end{figure*}

In training-free frameworks, LLMs rely on specialized tools to augment reasoning without fine-tuning. These tools are organized by the standard repair pipeline, progressing from bug reproduction, fault localization, and code search to patch generation, validation, and test generation(See Figure~\ref{fig:tool}).

\paragraph{Bug reproduction tools.} These tools automate the critical first step of debugging by generating executable scripts that trigger reported defects. Implementations typically leverage historical interaction data to adapt to repository-specific conventions~\cite{lin2024llmscontinuouslearnersimproving}, or employ finite state machines to govern behavior via multi-dimensional feedback, as in AEGIS~\cite{wang2025aegisagentbasedframeworkgeneral}.

\paragraph{Fault localization tools.} Once a bug is reproduced, these tools pinpoint suspicious code regions to narrow the search space. Common approaches include method-level Spectrum-Based Fault Localization (SBFL)~\cite{zhang2024autocoderoverautonomousprogramimprovement} and graph-based methods that construct code dependency graphs to trace fault propagation~\cite{li2025swedebatecompetitivemultiagentdebate}.

\paragraph{Code search tools.} These tools retrieve relevant dependency context after localization. Strategies range from interactive retrieval using BM25 or AST-based APIs~\cite{tao2024magisllmbasedmultiagentframework, yang2024sweagentagentcomputerinterfacesenable}, to graph-based global understanding via Knowledge Graphs and Language Server Protocols~\cite{ma2025alibabalingmaagentimprovingautomated, liu2024marscodeagentainativeautomated}, and dynamic managers that balance exploration breadth and depth~\cite{yu2025orcalocallmagentframework, jiang2025issuelocalizationllmdriveniterative}.

\paragraph{Patch generation tools.} These tools enhance LLM output quality through structured methodologies, including augmenting input context via specification inference~\cite{ruan2024specrovercodeintentextraction}, employing robust editing formats such as AutoDiff to bypass line-numbering failures~\cite{yang2024sweagentagentcomputerinterfacesenable, liu2024marscodeagentainativeautomated}, and employing ensemble selection mechanisms that filter candidates through regression testing~\cite{traeresearchteam2025traeagentllmbasedagent}.

\paragraph{Patch validation tools.} These tools confirm correctness and prevent regressions through external verification. Standard approaches include dynamic execution orchestration using sandboxed environments~\cite{zhang2024autocoderoverautonomousprogramimprovement, liu2024marscodeagentainativeautomated}, and static analysis mechanisms leveraging QA agents or Language Server Protocols for immediate diagnostic feedback~\cite{tao2024magisllmbasedmultiagentframework,liu2024marscodeagentainativeautomated}.

\paragraph{Test generation tools.} These tools generate reproduction test cases to validate intent and guide resolution. Systems typically employ feedback-driven iterative mechanisms that utilize error classification to synthesize failing tests reproducing reported defects, as in Otter~\cite{ahmed2025ottergeneratingtestsissues} and Issue2Test~\cite{nashid2025issue2testgeneratingreproducingtest}.

\paragraph{Other extensions.} Recent extensions focus on equipping agents with versatile tools to handle broader scenarios, including multimodal challenges. Strategies involve multimodal browsing and unified information access, which standardizes heterogeneous data into Markdown for seamless processing~\cite{soni2025codingagentsmultimodalbrowsing,lei2025infantagentnextmultimodalgeneralistagent}.

\subsubsection{Memory modules}
\label{sec:memory}

Memory integration empowers agents to transcend isolated problem-solving by accumulating historical context to guide future actions. Initial architectures focused on establishing hierarchical storage structures, such as segregating general knowledge from repository-specific details to mitigate rigidity~\cite{lin2024llmscontinuouslearnersimproving}, or archiving populations of agent variants to support open-ended evolution~\cite{zhang2025darwingodelmachineopenended}. To overcome the limitations of static prompting, \citet{mu2025experepairdualmemoryenhancedllmbased} incorporated dual-process cognitive architectures that synergize episodic records of concrete repairs with semantic layers of abstract insight, enabling dynamic retrieval based on current context. Current frontiers prioritize distilling transferable reasoning strategies, effectively shifting the paradigm from storing raw data to abstracting high-level policies from both successful and failed trajectories. This evolution allows agents to leverage multi-faceted experience banks to guide strategic search frameworks like MCTS~\cite{chen2025sweexpexperiencedrivensoftwareissue} and to prevent error repetition through generalized, rule-based learning~\cite{ouyang2025reasoningbankscalingagentselfevolving}.

\subsubsection{Inference-time scaling}
\label{sec:inference-time-scaling}
While specialized tools and memory systems enhance specific agent capabilities, relying solely on linear execution paths often limits the exploration of complex solution spaces. To address this, inference-time scaling has emerged as a critical paradigm to expand the search breadth and depth during problem-solving. To overcome the rigidity of sequential workflows, recent research focuses on enabling non-linear exploration via Monte Carlo Tree Search (MCTS), which facilitates flexible backtracking and qualitative feedback loops to prevent agents from stagnating in repetitive cycles~\cite{antoniades2025swesearchenhancingsoftwareagents}. Complementing this algorithmic shift, strategies for scaling computational resources deploy multiple independent state machines in parallel, maximizing solution coverage while amortizing the high costs of context identification without the need for model retraining~\cite{ehrlich2025codemonkeysscalingtesttimecompute}. Furthermore, advanced frameworks are now integrating memory-driven scaling, utilizing time-travel mechanisms to generate diverse experiences that serve not just to resolve the immediate task, but to distill generalizable reasoning strategies for long-term agent evolution~\cite{ouyang2025reasoningbankscalingagentselfevolving}.

\subsection{Training-based method}
\label{sec:training-methods}

Training-based methods encompass Supervised Fine-Tuning-based (SFT-based) methods and Reinforcement Learning-based (RL) methods, utilizing resources from Section~\ref{sec:training-datasets} to enhance the fundamental programming capabilities of LLMs.

\subsubsection{SFT-based method}
\label{sec:sft-methods}
Supervised Fine-Tuning (SFT) serves as the primary mechanism for grounding base models in software engineering protocols. Recent efforts to achieve robust domain adaptation focus on three key dimensions:
\noindent(1) \textit{Data Scaling}. Strategies increasingly prioritize the expansion of data scale and diversity via synthesized corpora. Frameworks employ iterative generation and filtering pipelines—augmented by automatic test generation or mid-training on billions of GitHub tokens—to comprehensively cover diverse repair scenarios~\cite{ma2025alibabalingmaagentimprovingautomated, wang2025swedevbuildingsoftwareengineering, yang2025kimidevagentlesstrainingskill}.
\noindent(2) \textit{Curriculum Learning}. Beyond raw volume, research emphasizes multi-stage curriculum learning. Models are refined through phased training sequences that progress from broad trajectory ingestion to strictly filtered, high-quality subsets or specialized tasks like localization and testing~\cite{He2025SWESwiss, rastogi2025devstralfinetuninglanguagemodels,liu2025contexttoolcontextmanagement}.
\noindent(3) \textit{Rejection Sampling}. To bridge the gap toward reinforcement learning, current methods employ rejection sampling pipelines. By fine-tuning exclusively on successful trajectories, these methods establish a strong baseline policy while simultaneously training verifiers to re-rank solutions at inference time~\cite{pan2025trainingsoftwareengineeringagents}. See Table~\ref{tab:swe-sft-methods} for detailed statistics on these SFT-trained models.

\subsubsection{RL-based method}
\label{sec:rl-methods}

Reinforcement learning optimizes issue resolution strategies through iterative interaction. This process hinges on the synergy of three core components: the algorithm for policy updates, reward design for guiding exploration, and the scaffold for managing environment rollouts. A statistical overview of recent models and their implementations across these dimensions is presented in Table~\ref{tab:swe-specialized-models} and Table~\ref{tab:general-models}.

\paragraph{Algorithm.}The optimization of agent behaviors leverages various policy gradient and alignment strategies to stabilize learning. We discuss three common algorithmic choices as follows. (1) \textit{Group Relative Policy Optimization (GRPO)}. A dominant approach employs GRPO, which enhances reasoning capabilities by normalizing advantages across group sampling without the heavy computational burden of a critic model~\cite{wei2025swerladvancingllmreasoning, sun2025scalinglonghorizonllmagent}. (2) \textit{Proximal Policy Optimization (PPO)}. Beyond group-based methods, some approaches utilize PPO for stable updates focused on subtasks~\cite{ma2025sorftissueresolvingsubtaskoriented}. (3) \textit{Direct Preference Optimization (DPO)}. Other work integrates MCTS with DPO to align complex, multi-step decision processes with high-quality preferred trajectories~\cite{zhang2025sealignalignmenttrainingsoftware}.

\paragraph{Reward design.}Effective feedback mechanisms typically incorporate both sparse outcome-based rewards and dense process-oriented signals. Most systems employ outcome reward models to provide terminal signals by utilizing strict metrics like patch similarity or detailed subtask verification ranging from localization to editing, thereby rigorously aligning outputs with ground truth~\cite{wei2025swerladvancingllmreasoning, ma2025sorftissueresolvingsubtaskoriented}. To mitigate the challenge of signal sparsity in long-horizon reasoning, researchers increasingly adopt process reward models and potential-based shaping techniques; these mechanisms provide dense, step-by-step feedback or token-level incentives, offering reward signals for intermediate behaviors such as context management and trajectory search throughout the reasoning process~\cite{zeng2025satorisweevolutionarytesttimescaling, da2025agentrlvrtrainingsoftwareengineering, sun2025scalinglonghorizonllmagent}.

\paragraph{Scaffold.} In the context of RL, a scaffold serves as the inference framework for rollouts. As statistics in Table \ref{tab:swe-specialized-models} indicate, OpenHands is the most prevalent scaffold, followed by workflow-based methods (notably Agentless and two-stage workflows). Environment-native frameworks like R2E-Gym and SWE-Gym are also frequently adopted due to their seamless alignment with training data.

\section{Analysis}
\label{sec:analysis}

Beyond developing new methodologies, a complementary line of research focuses on empirical analysis of existing data and methods, which provide critical insights into the limitations of current approaches and offer valuable perspectives for future research directions.

\subsection{Data analysis}
\label{sec:data-quality}
Recent scrutiny has exposed hidden benchmark defects, revealing that agent success rates are frequently inflated by solution leakage, ambiguous issue descriptions, and weak test suites that fail to catch incorrect patches~\cite{OpenAI2024swev, aleithan2024swebenchenhancedcodingbenchmark, wang_are_2025}. Recognizing that manual cleanup is too costly and inconsistent for large-scale datasets, the field is shifting toward automating validation workflows, utilizing model-based consensus mechanisms to reliably distinguish valid fixes from false positives without human intervention~\cite{oliva2025spiceautomatedswebenchlabeling}.

\subsection{Methods analysis}

\label{sec:method-analysis}
Research has shifted beyond measuring simple success rates to investigate the behavioral pathology of agents. A primary focus involves diagnosing internal reasoning failures, specifically examining the tendency of models to prioritize prolonged internal deliberation over necessary environmental interaction—a maladaptive pattern that leads to analysis paralysis and rogue actions~\cite{cuadron2025dangeroverthinkingexaminingreasoningaction}. Complementing this, efforts to manage the complexity of massive, 128k-token interaction logs have led to streamlining trajectory inspection, utilizing novel visual interfaces to transform cryptic output streams into navigable workflows for rapid error analysis~\cite{bula2025seaviewsoftwareengineeringagent}.

\section{Application}
The industrial deployment of software engineering AI has progressed from localized IDE assistance to fully autonomous systems capable of handling complex enterprise workflows. Due to space constraints, we provide a detailed discussion on these application scenarios in \S~\ref{sec:application-details}.

\section{Challenges and Opportunities}

\paragraph{High computational overhead.} In online RL, performing concurrent rollouts necessitates the simultaneous orchestration of numerous sandboxed containers, which incurs substantial storage footprints and computational costs. Similarly, verifying instances during data construction requires extensive parallel validation. This highlights the need for lightweight sandboxing and optimized resource scheduling.

\paragraph{Lack of efficiency-aware evaluation.} Current evaluations of issue resolution methods mainly focus on effectiveness metrics such as resolve rates while overlooking efficiency metrics like API costs and inference time. This oversight creates a biased domain where the computational and economic burdens of high-performing models are obscured. Consequently, future research must integrate both resolve rates and efficiency metrics into the evaluation framework to objectively reflect the comprehensive performance of issue resolution methods.

\paragraph{Limited Visually-Grounded Reasoning.} Multimodal tasks are rare in current benchmarks, hindering the evaluation of visually-dependent tasks such as frontend development and data visualization. Moreover, existing methods often simply flatten visuals into text, failing to capture the critical alignment between rendering and code. To address this, future research must prioritize constructing multimodal benchmarks and training specialized code-centric models~\cite{sun2025januscoderfoundationalvisualprogrammaticinterface}.

\paragraph{Safety risks in autonomous resolution.}

Recently, some agents have exhibited unsafe behaviors on coding tasks, including deleting a user's codebase~\cite{businessinsider2025replit} and cheating during evaluation\footnote{\url{https://github.com/SWE-bench/SWE-bench/issues/465}}. These failures motivate safer agent frameworks and more robust model safety alignment to prevent reward hacking in real deployments.

\paragraph{Lack of fine-grained rewards.}
Most RL methods for issue resolution still rely on outcome-level rewards, typically the binary test pass/fail signal. However, issue resolution requires multi-turn interaction with the environment, and an outcome reward makes credit assignment across action steps ambiguous. A promising direction is to design finer-grained process rewards  to provide denser supervision and improve policy optimization.

\paragraph{Data leakage and contamination.}
As benchmarks like SWE-Bench approach saturation, evaluation reliability is threatened by significant data leakage and quality control issues. Models may inadvertently memorize solutions due to unclear training cutoff dates~\cite{prathifkumar2025doesswebenchverifiedtestagent}, while the benchmarks themselves frequently suffer from invalid instances—including ambiguous descriptions, solution hints, and insufficient test coverage~\cite{OpenAI2024swev,mathews2025automatedsoftwareengineertrustworthy}. To restore trust, future frameworks must prioritize rigorous data curation and decontamination protocols to guarantee the validity of comparative assessments~\cite{zhu2025establishingbestpracticesbuilding}.

\paragraph{Lack of autonomous context management mechanisms.} Issue resolution tasks often require long-horizon, multi-turn interaction between the model and the code environment. This both raises API cost and degrades performance due to context rot~\cite{anthropic2024effective}. A promising solution is to construct an autonomous context management mechanism that proactively compresses and curates the model’s interaction history.

\paragraph{Insufficient patch validation and human review.} Since gold tests are unavailable in real-world development, relying solely on generation capability is insufficient. Future agents should incorporate intrinsic validation mechanisms, utilizing regression testing and dependency analysis to prevent feature regression. Additionally, to bridge the trust gap, research can prioritize human-centric interfaces, such as visual explanations and concise summaries, that assist developers in efficiently reviewing and accepting model-generated solutions.

\paragraph{Lack of universality across SWE domains.} While existing research predominantly focuses on the implementation and integration phases of the Software Development Life Cycle (SDLC), it often fails to address the comprehensive needs of the broader software engineering field. Future research should therefore broaden its scope to encompass diverse lifecycle stages—such as requirements analysis and architectural design—to develop more versatile automated software generation methods.

\section{Conclusion}
This paper presents a systematic survey of LLM-based issue resolution, synthesizing 175 publicly available papers and online resources in this rapidly evolving area. We introduce a tailored taxonomy that structures prior work along three core dimensions: data, methods, and analysis. Building on this structured view, we distill key challenges and outline promising directions toward more reliable, reproducible, and practical issue-resolution systems. To facilitate continued progress, we maintain an open-source repository that tracks relevant datasets, implementations, and new developments in the field.

\section{Limitations}

As the first dedicated survey on issue resolution, we prioritize high-level summaries over exhaustive details due to space constraints. Our search methodology relied on citation tracking (e.g., of SWE-bench) and snowballing; while thorough, this may overlook niche or nascent works. To address this rapid evolution, we commit to continuously updating our open-source repository.

\bibliography{main}

\newpage

\appendix

\section{Appendix}
\label{sec:appendix}

\subsection{Related work}
\paragraph{Code generation.} 
The application of LLMs in the programming domain has witnessed explosive growth. Early research focused primarily on function-level code generation, with benchmarks such as HumanEval~\cite{chen2021evaluatinglargelanguagemodels} serving as standard metrics. However, generic benchmarks often fail to capture the nuances of real-world development. To bridge this gap, recent initiatives~\cite{zheng2025generalperformancedomain,li2025feabenchbenchmarkevaluatingrepositorylevel,bogomolov2024longcodearenaset} have attempted to extend evaluation tasks to align more closely with realistic software development scenarios, revealing the limitations of general models in specialized domains. Concurrently, methods are also evolving to capture these broader contexts. While foundational approaches primarily relied on SFT~\cite{Liu2024MFTCoder} or standard retrieval-augmented generation~\cite{wu2024repoformer,wang2025coderag}, RL-based methods emerged as a pivotal direction for handling complex coding tasks~\cite{wang2024rlcoderreinforcementlearningrepositorylevel}.

\paragraph{Automated software generation.} The primary goal of this task is to autonomously construct complete and executable software systems starting from high-level natural language requirements. Unlike code completion, it necessitates covering the SDLC, including requirement analysis, system design, coding, and testing. To address the complexity and potential logic inconsistencies in this process, state-of-the-art frameworks like ChatDev~\cite{qian2024chatdevcommunicativeagentssoftware} and MetaGPT~\cite{hong2024metagptmetaprogrammingmultiagent} leverage multi-agent collaboration, simulating human development teams to decompose complex tasks into streamlined and verifiable workflows. Rencently ~\cite{luo2025rpgrepositoryplanninggraph} achieve repository-level generation from scratch. It introduces the Repository Planning Graph (RPG), which encodes file structures and data flows into a unified graph, effectively replacing free-form natural language with an explicit blueprint for consistent long-horizon planning.

\paragraph{Automated software maintenance.} Issue resolution is intrinsically linked to the broader domain of automated software maintenance. Methodologies established in this field are frequently encapsulated as callable tools to augment the capabilities of LLMs in software development tasks. Key techniques utilized to enhance LLM performance include bug reproduction approaches~\cite{khatib2026assertflipreproducingbugsinversion,kitsios2025automatedgenerationissuereproducingtests}, fault localization approaches such as SBFL~\cite{wong2016sfl,zhou2012whereshouldthebug,qi2022dreamloc,hu2014effectivebugtriage,Chakraborty2025blaze,chang2025bridgingbuglocalizationissue,zhang2025hierarchical}, code search approaches~\cite{zhang2025benchmarklocalizingcodenoncode}, and test generation approaches~\cite{ahmed2024tddbenchverifiedllmsgenerate,Mündler2024swtbench,han2025tdflowagenticworkflowstest}. These tools provide agents with precise error locations, relevant code context, and verification mechanisms necessary for effective resolution. Furthermore, recent research has expanded the scope of automated software maintenance by utilizing issue resolution data to construct dialogue datasets that capture real-world human-computer collaboration~\cite{garg2025savingswebenchbenchmarkmutation}, and critically examining the security risks inherent in agent-generated code through vulnerability benchmarking~\cite{zhao2025vibecodingsafebenchmarking}.

\paragraph{Automated environment setup.} Recent initiatives focus on automating the configuration of runtime environments for entire repositories~\cite{eliseeva2025envbench, kovrigin2025piperondeviceenvironmentsetup, vergopoulos2025automated}. This capability develops in parallel with data construction for issue resolution.

\paragraph{Related surveys.} Existing surveys primarily focus on code generation~\cite{jiang2024survey,wang2025rag} or other tasks within the software engineering domain~\cite{huang2023surveyautomatedprogramrepair,zhang2023surveyoflearningbasedAPR,guo2025comprehensivesurveybenchmarkssolutions,wang2024agentssoftwareengineeringsurvey}. This paper bridges this gap by offering the first systematic survey dedicated to the entire spectrum of issue resolution, ranging from non-agent approaches to the latest agentic advancements.

\subsection{Detailed discussions on background}
\label{sec:background-details}

The issue resolution task aims to automatically resolve reported issues. As illustrated in Figure~\ref{fig:task} , it is formally defined by an instance represented as a triple $\mathcal{\gI}=(\gD,\gC,\gT)$, where these components map directly to the benchmark's metadata structure:
\begin{itemize}
    \item $\gD$: Issue description, which is the original GitHub issue text description detailing the bug or feature request.
    \item $\gC$: Codebase, the collection of source files at the specific commit state before the issue was resolved. This state is precisely defined by the repository identifier (owner/repo) and a base commit hash, requiring the evaluation environment to perform a git clone and subsequent git checkout to establish the exact pre-fix version.
    \item $\gT$: Tests, the complete set of unit and system tests associated with the issue, derived from the original developer's test patch. This set is explicitly categorized into two subsets: $T_{\mathrm{fail}\rightarrow\mathrm{pass}}$, containing tests that fail on $C$ and must pass after the patch (verifying the fix), and $T_{\mathrm{pass}\rightarrow\mathrm{pass}}$, containing tests that must pass both before and after (ensuring no regression). 
\end{itemize}
Given only the inputs $(\gD, \gC)$, the model proposes an edit to resolve the issue. The model's output is a Patch $\gP$ (represented as a patch file). (The target solution is the gold patch provided in the dataset, which is not revealed to the model). The patch $\gP$ must be successfully applied to $\gC$ using a standard patch application utility. The codebase resulting from this application is $\gC'$. Crucially, the tests $\gT$ are not revealed to the model during this process.

\paragraph{Evaluation.} To evaluate a proposed solution (patch $\gP$), it is first applied to the original codebase $\gC$ to obtain $\gC'$. The repository's test suite $\gT$ is then executed on $\gC'$. The evaluation framework captures the output in a test log, which is subsequently processed by a log parser to determine the test results (e.g., pass or fail). Based on the parsed results, the solution is scored: it is considered successful if the patch applies correctly and all tests in $\gT$ pass. The final benchmark metric is the Resolve Rate, defined as the percentage of tasks that are successfully resolved.

\label{sec:data-construction-appendix}
\begin{table*}[t]
  \centering
  \scriptsize
  \renewcommand{\arraystretch}{1.1}
  
  \begin{tabular}{l >{\RaggedRight}m{3cm} c c r c >{\centering\arraybackslash}m{1.5cm}}
    \toprule
    \textbf{Dataset} & \textbf{Language} & \textbf{Multimodal} & \textbf{Repos} & \textbf{Amount} & \textbf{Environment} & \textbf{Link} \\
    \midrule

    \rowcolor[rgb]{0.88,0.95,1} \multicolumn{7}{c}{\textbf{Single-PL Datasets}} \\
    \addlinespace
    SWE-Fixer & Python & {\color[rgb]{1,0,0}\ding{55}} & 856 & 115,406 & {\color[rgb]{1,0,0}\ding{55}} & \ghlink{https://github.com/InternLM/SWE-Fixer} \hflink{https://huggingface.co/datasets/internlm/SWE-Fixer-Train-110K} \hflink{https://huggingface.co/datasets/internlm/SWE-Fixer-Eval} \\
    SWE-smith & Python & {\color[rgb]{1,0,0}\ding{55}} & 128 & 50k & {\color[rgb]{0,0.5,0}\ding{51}} & \ghlink{https://github.com/SWE-bench/SWE-smith} \hflink{https://huggingface.co/datasets/SWE-bench/SWE-smith} \\
    SWE-Lego & Python & {\color[rgb]{1,0,0}\ding{55}} & 3,251 & 32,119 & {\color[rgb]{0,0.5,0}\ding{51}} & \ghlink{https://github.com/SWE-Lego/SWE-Lego} \hflink{https://huggingface.co/SWE-Lego/datasets} \\
    SWE-rebench & Python & {\color[rgb]{1,0,0}\ding{55}} & 3,468 & 21,336 & {\color[rgb]{0,0.5,0}\ding{51}} & \ghlink{https://github.com/SWE-rebench/SWE-bench-fork} \hflink{https://huggingface.co/datasets/nebius/SWE-rebench} \\
    SWE-bench-train & Python & {\color[rgb]{1,0,0}\ding{55}} & 37 & 19k & {\color[rgb]{1,0,0}\ding{55}} & \ghlink{https://github.com/SWE-bench/SWE-bench} \hflink{https://huggingface.co/datasets/princeton-nlp/SWE-bench/viewer/default/train} \\
    SWE-Flow & Python & {\color[rgb]{1,0,0}\ding{55}} & 74 & 18,081 & {\color[rgb]{0,0.5,0}\ding{51}} & \ghlink{https://github.com/Hambaobao/SWE-Flow} \\
    Skywork-SWE & Python & {\color[rgb]{1,0,0}\ding{55}} & 2,531 & 10,169 & {\color[rgb]{0,0.5,0}\ding{51}} & / \\
    R2E-Gym & Python & {\color[rgb]{1,0,0}\ding{55}} & 10 & 8,135 & {\color[rgb]{0,0.5,0}\ding{51}} & \ghlink{https://github.com/R2E-Gym/R2E-Gym} \hflink{https://huggingface.co/R2E-Gym/datasets} \\
    RepoForge & Python & {\color[rgb]{1,0,0}\ding{55}} & / & 7.3k & {\color[rgb]{0,0.5,0}\ding{51}} & / \\
    SWE-bench-extra & Python & {\color[rgb]{1,0,0}\ding{55}} & 2k & 6.38k & {\color[rgb]{0,0.5,0}\ding{51}} &  \hflink{https://huggingface.co/datasets/nebius/SWE-bench-extra} \\
    SWE-Gym & Python & {\color[rgb]{1,0,0}\ding{55}} & 11 & 2,438 & {\color[rgb]{0,0.5,0}\ding{51}} & \ghlink{https://github.com/SWE-Gym/SWE-Gym} \hflink{https://huggingface.co/SWE-Gym/datasets} \\
    SWE-bench & Python & {\color[rgb]{1,0,0}\ding{55}} & 12 & 2,294 & {\color[rgb]{0,0.5,0}\ding{51}} & \ghlink{https://github.com/SWE-bench/SWE-bench} \hflink{https://huggingface.co/datasets/princeton-nlp/SWE-bench} \\
    SWE-bench-java & Java & {\color[rgb]{1,0,0}\ding{55}} & 19 & 1,797 & {\color[rgb]{0,0.5,0}\ding{51}} & \ghlink{https://github.com/multi-swe-bench/multi-swe-bench-env} \hflink{https://huggingface.co/datasets/Daoguang/Multi-SWE-bench} \\
    FEA-bench & Python & {\color[rgb]{1,0,0}\ding{55}} & 83 & 1,401 & {\color[rgb]{0,0.5,0}\ding{51}} & \ghlink{https://github.com/microsoft/FEA-Bench/}\hflink{https://huggingface.co/datasets/microsoft/FEA-Bench} \\
    SWE-bench-Live & Python & {\color[rgb]{1,0,0}\ding{55}} & 164 & 1,565 & {\color[rgb]{0,0.5,0}\ding{51}} & \ghlink{https://github.com/microsoft/SWE-bench-Live} \hflink{https://huggingface.co/datasets/SWE-bench-Live/SWE-bench-Live} \\
    Loc-Bench & Python & {\color[rgb]{1,0,0}\ding{55}} & / & 560 & {\color[rgb]{1,0,0}\ding{55}} & \ghlink{https://github.com/gersteinlab/LocAgent} \hflink{https://huggingface.co/datasets/czlll/Loc-Bench_V1} \\
    SWE-bench Verified & Python & {\color[rgb]{1,0,0}\ding{55}} & / & 500 & {\color[rgb]{0,0.5,0}\ding{51}} & \ghlink{https://github.com/SWE-bench/SWE-bench} \hflink{https://huggingface.co/datasets/princeton-nlp/SWE-bench_Verified} \\
    SWE-bench Lite & Python & {\color[rgb]{1,0,0}\ding{55}} & 12 & 300 & {\color[rgb]{0,0.5,0}\ding{51}} & \ghlink{https://github.com/SWE-bench/SWE-bench} \hflink{https://huggingface.co/datasets/princeton-nlp/SWE-bench_Lite} \\
    SWE-MERA & Python & {\color[rgb]{1,0,0}\ding{55}} &  200 & 300 & {\color[rgb]{0,0.5,0}\ding{51}} & \ghlink{https://github.com/MERA-Evaluation/SWE-MERA-submissions} \hflink{https://huggingface.co/datasets/MERA-evaluation/SWE-MERA}\\
    SWE-Bench-CL & Python & {\color[rgb]{1,0,0}\ding{55}} & 8 & 273 & {\color[rgb]{0,0.5,0}\ding{51}} & \ghlink{https://github.com/thomasjoshi/agents-never-forget} \\
    SWE-Sharp-Bench & C\# & {\color[rgb]{1,0,0}\ding{55}} & 17 & 150 & {\color[rgb]{0,0.5,0}\ding{51}} & \ghlink{https://github.com/microsoft/prose/tree/main/misc/SWE-Sharp-Bench} \hflink{https://huggingface.co/datasets/microsoft/SWE-Sharp-Bench}\\
    SWE-Perf & Python & {\color[rgb]{1,0,0}\ding{55}} & 12 & 140 & {\color[rgb]{0,0.5,0}\ding{51}} & \ghlink{https://github.com/SWE-Perf/swe-perf} \hflink{https://huggingface.co/datasets/SWE-Perf/SWE-Perf} \\
    Visual SWE-bench & Python & {\color[rgb]{0,0.5,0}\ding{51}} & 11 & 133 & {\color[rgb]{0,0.5,0}\ding{51}} & \ghlink{https://github.com/luolin101/CodeV} \hflink{https://huggingface.co/datasets/luolin101/Visual-SWE-bench} \\
    SWE-EVO & Python & {\color[rgb]{1,0,0}\ding{55}} & 7 & 48 & {\color[rgb]{0,0.5,0}\ding{51}} & \ghlink{https://github.com/bdqnghi/SWE-EVO} \\

    \addlinespace
    
    \rowcolor[rgb]{0.88,0.95,1} \multicolumn{7}{c}{\textbf{Multi-PL Datasets}} \\
    \addlinespace
    SWE-Mirror & Python, Rust, Go & {\color[rgb]{1,0,0}\ding{55}} & 40 & 60k & {\color[rgb]{0,0.5,0}\ding{51}} & / \\
    Multi-SWE-bench & Java, JS, TS, Go, Rust, C, C++ & {\color[rgb]{1,0,0}\ding{55}} & 76 & 4,723 & {\color[rgb]{0,0.5,0}\ding{51}} & \ghlink{https://github.com/multi-swe-bench/multi-swe-bench} \hflink{https://huggingface.co/datasets/ByteDance-Seed/Multi-SWE-bench} \\
    Swing-Bench & Python, Go, C++, Rust & {\color[rgb]{1,0,0}\ding{55}} & 400 & 2300 & {\color[rgb]{0,0.5,0}\ding{51}} & / \\
    SWE-PolyBench & Python, Java, JS, TS & {\color[rgb]{1,0,0}\ding{55}} & 21 & 2,110 & {\color[rgb]{0,0.5,0}\ding{51}} & \ghlink{https://github.com/amazon-science/SWE-PolyBench} \hflink{https://huggingface.co/datasets/AmazonScience/SWE-PolyBench} \hflink{https://huggingface.co/datasets/Sellopale/SWE-PolyBench_500}\\
    SWE-Compass & Python, JS, TS, Java, C, C++, Go, Rust, Kotlin, C\# & {\color[rgb]{1,0,0}\ding{55}} & / & 2,000 & {\color[rgb]{0,0.5,0}\ding{51}} & \ghlink{https://github.com/kwaipilot/SWE-Compass/} \hflink{https://huggingface.co/datasets/Kwaipilot/SWE-Compass} \\
    SWE-Bench Pro & Python, Go, TS  & {\color[rgb]{1,0,0}\ding{55}} & 41 & 1,865 & {\color[rgb]{0,0.5,0}\ding{51}} & \ghlink{https://github.com/scaleapi/SWE-bench_Pro-os} \hflink{https://huggingface.co/datasets/ScaleAI/SWE-bench_Pro} \\
    SWE-bench++ & Python, Go, TS, JS, Ruby, PHP, Java, Rust, C++, C\#, C & {\color[rgb]{1,0,0}\ding{55}} & 3,971 & 1,782 & {\color[rgb]{0,0.5,0}\ding{51}} & \ghlink{https://github.com/TuringEnterprises/SWE-Bench-plus-plus} \hflink{https://huggingface.co/datasets/TuringEnterprises/SWE-Bench-plus-plus} \\
    SWE-Lancer & JS, TS & {\color[rgb]{1,0,0}\ding{55}} & / & 1,488 & {\color[rgb]{0,0.5,0}\ding{51}} & \ghlink{https://github.com/openai/frontier-evals} \\
    OmniGIRL & Python, TS, Java, JS & {\color[rgb]{0,0.5,0}\ding{51}} & 15 & 959 & {\color[rgb]{0,0.5,0}\ding{51}} & \ghlink{https://github.com/deepsoftwareanalytics/omnigirl} \hflink{https://huggingface.co/datasets/Deep-Software-Analytics/OmniGIRL} \\
    SWE-bench Multimodal & JS, TS, HTML, CSS & {\color[rgb]{0,0.5,0}\ding{51}} & 17 & 619 & {\color[rgb]{0,0.5,0}\ding{51}} & \ghlink{https://github.com/SWE-bench/SWE-bench} \hflink{https://huggingface.co/datasets/SWE-bench/SWE-bench_Multimodal} \\
    SWE-fficiency & Python, Cython & {\color[rgb]{1,0,0}\ding{55}} & 9 & 498 & {\color[rgb]{0,0.5,0}\ding{51}} & \ghlink{https://github.com/swefficiency/swefficiency-site} \\
    SWE-Factory & Python, Java, JS, TS & {\color[rgb]{1,0,0}\ding{55}} & 12 & 430 & {\color[rgb]{0,0.5,0}\ding{51}} & \ghlink{https://github.com/DeepSoftwareAnalytics/swe-factory} \hflink{https://huggingface.co/SWE-Factory} \\
    SWE-bench-Live-MultiLang \& Windows & Python, JS, TS, C, C++, C\#,  Java, Go, Rust & {\color[rgb]{1,0,0}\ding{55}} & 238 & 418 & {\color[rgb]{0,0.5,0}\ding{51}} & \ghlink{https://github.com/microsoft/SWE-bench-Live} \hflink{https://huggingface.co/datasets/SWE-bench-Live/MultiLang} \hflink{https://huggingface.co/datasets/SWE-bench-Live/Windows} \\
    SWE-bench Multilingual & C, C++, Go, Java, JS, TS, Rust, Python, Ruby, PHP & {\color[rgb]{1,0,0}\ding{55}} & 42 & 300 & {\color[rgb]{0,0.5,0}\ding{51}} & \ghlink{https://github.com/SWE-bench/SWE-bench} \hflink{https://huggingface.co/datasets/SWE-bench/SWE-bench_Multilingual} \\
    SWE-InfraBench & Python, TS & {\color[rgb]{1,0,0}\ding{55}} & / & 100 & {\color[rgb]{0,0.5,0}\ding{51}} & / \\

    \bottomrule
  \end{tabular}
      \caption{A comprehensive survey and statistical overview of issue resolution datasets. We categorize these datasets based on programming language, modality support, source repositories, data scale (Amount), and the availability of reproducible execution environments.}  
    \label{tab:dataset-survey}

\end{table*}
\begin{table*}[t]
  \centering
  \small
  \renewcommand{\arraystretch}{1.1}

  \begin{tabular}{l l r r >{\centering\arraybackslash}p{1.5cm}}
    \toprule
    \textbf{Dataset} & \textbf{Language} & \textbf{Repos} & \textbf{Amount} & \textbf{Link} \\
    \midrule
    R2E-Gym & Python & 10 & 3,321 & \ghlink{https://github.com/R2E-Gym/R2E-Gym} \hflink{https://huggingface.co/datasets/R2E-Gym/R2EGym-SFT-Trajectories} \\
    SWE-Gym & Python & 11 & 491 & \ghlink{https://github.com/SWE-Gym/SWE-Gym} \hflink{https://huggingface.co/datasets/SWE-Gym/OpenHands-SFT-Trajectories} \\
    SWE-Synth & Python & 11 & 3,018 & \ghlink{https://github.com/FSoft-AI4Code/SWE-Synth} \hflink{https://huggingface.co/datasets/swesynth/SWE-Synth_Moatless-SFT-Trajectories} \\
    SWE-Fixer & Python & 856 & 69,752 & \ghlink{https://github.com/InternLM/SWE-Fixer} \hflink{https://huggingface.co/datasets/internlm/SWE-Fixer-Train-Editing-CoT-70K} \\
    SWE-Factory & Python & 10 & 2,809 & \ghlink{https://github.com/DeepSoftwareAnalytics/swe-factory} \hflink{https://huggingface.co/datasets/SWE-Factory/DeepSWE-Agent-Kimi-K2-Trajectories-2.8K} \\
    SWE-rebench & Python & 1,823 & 67,074 & \hflink{https://huggingface.co/datasets/nebius/SWE-rebench-openhands-trajectories} \\
    SWE-Lego & Python & 3251 &  14.6k & \ghlink{https://github.com/SWE-Lego/SWE-Lego}\\    
    \bottomrule
  \end{tabular}
    \caption{A survey of trajectory datasets used for agent training or analysis. We list the programming language, number of source repositories, and total trajectories for each dataset.}
  \label{tab:trajectory-datasets}

\end{table*}

\paragraph{Data Construction.}This process is structured into four main stages.

First, (1) \textit{Repo Selection and Data Scraping} involves collecting a large set of PRs from popular, well-maintained open-source repositories (e.g., 12 popular Python repositories for the original SWE-bench).

Second, (2) \textit{Attribute-based Filtering} narrows down the candidates, selecting only merged PRs that are documented to resolve a specific GitHub issue and that make modifications to the repository's test files (indicating that tests were contributed).

Third, (3) \textit{Execution-based Filtering} is a critical stage that ensures tasks are reproducible and valid. To guarantee reliable environment construction, recent works increasingly leverage CI/CD configurations—specifically GitHub Actions workflows found under \texttt{.github/workflows/}—as the ground truth for dependency management. For instance, Multi-SWE-bench~\cite{zan2025multiswebenchmultilingualbenchmarkissue} extracts build steps from these workflows to create isolated Dockerized environments. Similarly, SWE-Sharp-Bench~\cite{mhatre2025swesharpbenchreproduciblebenchmarkc} confirms build viability by executing local GitHub Actions workflows, ensuring the repository successfully builds and passes tests at the latest commit. By filtering out instances that fail these automated installation or runtime checks, this stage establishes a stable foundation. Finally, it verifies the PR's corrective nature by executing the test suite before and after applying the patch, retaining only instances that demonstrate at least one clear Fail-to-Pass (F2P) test transition.

Finally, (4) \textit{Manual Verification} is performed to ensure the quality and usability of the filtered tasks. This step often involves human inspection to check for the clarity of the problem description (issue), the self-contained nature of the task, and its overall suitability for the benchmark.

While this initial pipeline was effective for creating a static dataset (e.g., the 2,294 SWE-bench instances), its reliance on a complex, manual-heavy environment setup (especially in stages 3 and 4) and its susceptibility to data contamination limited its long-term scalability and dynamism. Consequently, subsequent research has focused on enhancing these data methods, leading to two major axes of improvement: Data Collection for building more dynamic benchmarks, and Data Synthesis for creating high-quality synthetic datasets for training large language models.

\subsection{Detailed discussions on data}
\label{sec:data-detail}

\begin{table*}[t]
  \centering
  \scriptsize
  \renewcommand{\arraystretch}{1.3} 
  \setlength{\tabcolsep}{2pt}       
  
  \begin{tabular}{
      >{\RaggedRight\bfseries}m{2.8cm} 
      >{\RaggedRight}m{2.6cm}          
      c                                
      c                                
      >{\RaggedRight}m{3.0cm}          
      c                                
      c c c                            
  }
  
    \toprule
    \textbf{Model Name} & \textbf{Base Model} & \textbf{Size} & \textbf{Arch.} & \textbf{Training Scaffold} & \textbf{Res.(\%)} & \textbf{Code} & \textbf{Data} & \textbf{Model} \\
    \midrule

    SWE-rebench-openhands-Qwen3-235B-A22B & 
    Qwen3-235B-A22B & 
    235B-A22B & 
    MoE & 
    OpenHands & 
    59.9 & 
    / & 
    \hflink{https://huggingface.co/datasets/nebius/SWE-rebench-openhands-trajectories}& 
    \hflink{https://huggingface.co/nebius/SWE-rebench-openhands-Qwen3-235B-A22B} \\
    \addlinespace

    SWE-Lego-Qwen3-32B & 
    Qwen3-32B & 
    32B & 
    Dense & 
    OpenHands & 
    57.6 & 
    \ghlink{https://github.com/SWE-Lego/SWE-Lego} & 
    \hflink{https://huggingface.co/SWE-Lego/datasets}& 
    \hflink{https://huggingface.co/SWE-Lego/SWE-Lego-Qwen3-32B} \\
    \addlinespace

    SWE-rebench-openhands-Qwen3-30B-A3B & 
    Qwen3-30B-A3B & 
    30B-A3B & 
    MoE & 
    OpenHands & 
    49.7 & 
    / & 
    \hflink{https://huggingface.co/datasets/nebius/SWE-rebench-openhands-trajectories}& 
    \hflink{https://huggingface.co/nebius/SWE-rebench-openhands-Qwen3-30B-A3B} \\
    \addlinespace

    Devstral & 
    Mistral Small 3 & 
    22B & 
    Dense & 
    OpenHands & 
    46.8 & 
    / & 
    \bloglink{https://mistral.ai/news/devstral} & 
    \hflink{https://huggingface.co/mistralai/Devstral-Small-2507} \\
    \addlinespace
    
    Co-PatcheR & 
    Qwen2.5-Coder-14B & 
    3$\times$14B & 
    Dense & 
    PatchPilot-mini & 
    46.0 & 
    \ghlink{https://github.com/ucsb-mlsec/Co-PatcheR} & 
    / & 
    \hflink{https://huggingface.co/collections/UCSB-SURFI/co-patcher} \\
    \addlinespace
    
    SWE-Swiss-32B & 
    Qwen2.5-32B-Instruct & 
    32B & 
    Dense & 
    Agentless & 
    45.0 & 
    \ghlink{https://github.com/zhenyuhe00/SWE-Swiss} & 
    \hflink{https://huggingface.co/SWE-Swiss/datasets} & 
    \hflink{https://huggingface.co/SWE-Swiss/models} \\
    \addlinespace

    SWE-Lego-Qwen3-8B & 
    Qwen3-8B & 
    8B & 
    Dense & 
    OpenHands & 
    44.4 & 
    \ghlink{https://github.com/SWE-Lego/SWE-Lego} & 
    \hflink{https://huggingface.co/SWE-Lego/datasets}& 
    \hflink{https://huggingface.co/SWE-Lego/SWE-Lego-Qwen3-8B} \\
    \addlinespace

    Lingma SWE-GPT & 
    Qwen2.5-72B-Instruct & 
    72B & 
    Dense & 
    SWESynInfer & 
    30.2 & 
    \ghlink{https://github.com/LingmaTongyi/Lingma-SWE-GPT} & 
    / & 
    / \\
    \addlinespace

    SWE-Gym-Qwen-32B & 
    Qwen2.5-Coder-32B & 
    32B & 
    Dense & 
    OpenHands, MoatlessTools & 
    20.6 & 
    \ghlink{https://github.com/SWE-Gym/SWE-Gym} & 
    / & 
    \hflink{https://huggingface.co/SWE-Gym} \\
    \addlinespace

    Lingma SWE-GPT & 
    Qwen2.5-Coder-7B & 
    7B & 
    Dense & 
    SWESynInfer & 
    18.2 & 
    \ghlink{https://github.com/LingmaTongyi/Lingma-SWE-GPT} & 
    / & 
    / \\
    \addlinespace

    SWE-Gym-Qwen-14B & 
    Qwen2.5-Coder-14B & 
    14B & 
    Dense & 
    OpenHands, MoatlessTools & 
    16.4 & 
    \ghlink{https://github.com/SWE-Gym/SWE-Gym} & 
    / & 
    \hflink{https://huggingface.co/SWE-Gym} \\
    \addlinespace
    
    SWE-Gym-Qwen-7B & 
    Qwen2.5-Coder-7B & 
    7B & 
    Dense & 
    OpenHands, MoatlessTools & 
    10.6 & 
    \ghlink{https://github.com/SWE-Gym/SWE-Gym} & 
    / & 
    \hflink{https://huggingface.co/SWE-Gym} \\
    
    \bottomrule
  \end{tabular}
    \caption{Overview of SFT-based methods for issue resolution. This table categorizes models by their base architecture and training scaffold (Sorted by Performance).}
  \label{tab:swe-sft-methods}

\end{table*}
\begin{table*}[t]
  \centering
  \scriptsize
  \renewcommand{\arraystretch}{1.2}
  \setlength{\tabcolsep}{2pt}

  \begin{tabular}{
      >{\RaggedRight\bfseries}p{2.5cm} 
      >{\RaggedRight}p{2.5cm}          
      c                                
      c                                
      >{\RaggedRight}p{2.7cm}          
      c                                
      c                                
      c c c                            
  }
    \toprule
    \textbf{Model Name} & \textbf{Base Model} & \textbf{Size} & \textbf{Arch.} & \textbf{Train. Scaffold} & \textbf{Reward} & \textbf{Res.(\%)} & \textbf{Code} & \textbf{Data} & \textbf{Model} \\
    \midrule
    
    \multicolumn{10}{c}{\cellcolor[rgb]{0.88,0.95,1}\textbf{560B Models (MoE)}} \\
    \addlinespace
    LongCat-Flash-Think & LongCatFlash-Base & 560B-A27B & MoE & R2E-Gym & Outcome & 60.4 & \ghlink{https://github.com/meituan-longcat/LongCat-Flash-Thinking} & / & \hflink{https://huggingface.co/meituan-longcat/LongCat-Flash-Thinking} \\
    \addlinespace
    
    \multicolumn{10}{c}{\cellcolor[rgb]{0.88,0.95,1}\textbf{72B Models}} \\
    \addlinespace
    Kimi-Dev & Qwen 2.5-72B-Base & 72B & Dense & BugFixer + TestWriter & Outcome & 60.4 & \ghlink{https://github.com/MoonshotAI/Kimi-Dev} & / & \hflink{https://huggingface.co/moonshotai/Kimi-Dev-72B} \\
    Multi-turn RL(Nebius) & Qwen2.5-72B-Instruct & 72B & Dense & SWE-agent & Outcome & 39.0 & / & / & / \\
    Agent-RLVR-RM-72B & Qwen2.5-Coder-72B & 72B & Dense & Localization + Repair & Outcome & 27.8 & / & / & / \\
    Agent-RLVR-72B & Qwen2.5-Coder-72B & 72B & Dense & Localization + Repair & Outcome & 22.4 & / & / & / \\
    \addlinespace
    
    \multicolumn{10}{c}{\cellcolor[rgb]{0.88,0.95,1}\textbf{70B Models}} \\
    \addlinespace
    SWE-RL & Llama-3.3-70B-Instruct & 70B & Dense & Agentless-mini & Outcome & 41.0 & \ghlink{https://github.com/facebookresearch/swe-rl} & / & / \\
    \addlinespace
    
    \multicolumn{10}{c}{\cellcolor[rgb]{0.88,0.95,1}\textbf{36B Models}} \\
    \addlinespace
    FoldAgent & Seed-OSS-36B-Instruct & 36B & Dense & FoldAgent & Process & 58.0 & \ghlink{https://github.com/sunnweiwei/FoldAgent} & \bloglink{https://drive.google.com/file/u/0/d/1aX5xXAN5R-gLKd8A0AY-troxXJRawyAM/view?usp=sharing\&pli=1} & / \\
    \addlinespace
    
    \multicolumn{10}{c}{\cellcolor[rgb]{0.88,0.95,1}\textbf{32B Models}} \\
    \addlinespace
    OpenHands Critic & Qwen2.5-Coder-32B & 32B & Dense & SWE-Gym & / & 66.4 & \ghlink{https://github.com/All-Hands-AI/OpenHands} & / & \hflink{https://huggingface.co/OpenHands/openhands-critic-32b-exp-20250417} \\
    KAT-Dev-32B & Qwen3-32B & 32B & Dense & / & / & 62.4 & / & / & \hflink{https://huggingface.co/Kwaipilot/KAT-Dev} \\
    SWE-Swiss-32B & Qwen2.5-32B-Instruct & 32B & Dense & / & Outcome & 60.2 & \ghlink{https://github.com/zhenyuhe00/SWE-Swiss} & \hflink{https://huggingface.co/SWE-Swiss/datasets} & \hflink{https://huggingface.co/SWE-Swiss/models} \\
    SeamlessFlow-32B & Qwen3-32B & 32B & Dense & SWE-agent & Outcome & 45.8 & \ghlink{https://github.com/Chojikun/seamlessflow} & / & / \\
    DeepSWE & Qwen3-32B & 32B & Dense & R2E-Gym & Outcome & 42.2 & \ghlink{https://github.com/agentica-project/rllm} & \hflink{https://huggingface.co/datasets/R2E-Gym/R2E-Gym-Subset} & \hflink{https://huggingface.co/agentica-org/DeepSWE-Preview} \\
    SA-SWE-32B & / & 32B & Dense & SkyRL-Agent & / & 39.4 & / & / & / \\
    OpenHands LM v0.1 & Qwen2.5-Coder-32B & 32B & Dense & SWE-Gym & / & 37.2 & \ghlink{https://github.com/All-Hands-AI/OpenHands} & / & \hflink{https://huggingface.co/OpenHands/openhands-lm-32b-v0.1} \\
    SWE-Dev-32B & Qwen2.5-Coder-32B & 32B & Dense & OpenHands & Outcome & 36.6 & \ghlink{https://github.com/THUDM/SWE-Dev} & / & \hflink{https://huggingface.co/zai-org/SWE-Dev-32B} \\
    Satori-SWE & Qwen2.5-Coder-32B & 32B & Dense & Retriever + Code editor & Outcome & 35.8 & \ghlink{https://github.com/satori-reasoning/Satori-SWE} & \hflink{https://huggingface.co/Satori-reasoning} & \hflink{https://huggingface.co/Satori-reasoning} \\
    SoRFT-32B & Qwen2.5-Coder-32B & 32B & Dense & Agentless & Outcome & 30.8 & / & / & / \\
    Agent-RLVR-32B & Qwen2.5-Coder-32B & 32B & Dense & Localization + Repair & Outcome & 21.6 & / & / & / \\
    \addlinespace
    
    \multicolumn{10}{c}{\cellcolor[rgb]{0.88,0.95,1}\textbf{14B Models}} \\
    \addlinespace
    Agent-RLVR-14B & Qwen2.5-Coder-14B & 14B & Dense & Localization + Repair & Outcome & 18.0 & / & / & / \\
    SEAlign-14B & Qwen2.5-Coder-14B & 14B & Dense & OpenHands & Process & 17.7 & / & / & / \\
    \addlinespace
    
    \multicolumn{10}{c}{\cellcolor[rgb]{0.88,0.95,1}\textbf{9B Models}} \\
    \addlinespace
    SWE-Dev-9B & GLM-4-9B & 9B & Dense & OpenHands & Outcome & 13.6 & \ghlink{https://github.com/THUDM/SWE-Dev} & / & \hflink{https://huggingface.co/zai-org/SWE-Dev-9B} \\
    \addlinespace
    
    \multicolumn{10}{c}{\cellcolor[rgb]{0.88,0.95,1}\textbf{8B Models}} \\
    \addlinespace
    SeamlessFlow-8B & Qwen3-8B & 8B & Dense & SWE-agent & Outcome & 27.4 & \ghlink{https://github.com/Chojikun/seamlessflow} & / & / \\
    SWE-Dev-8B & Llama-3.1-8B & 8B & Dense & OpenHands & Outcome & 18.0 & \ghlink{https://github.com/THUDM/SWE-Dev} & / & \hflink{https://huggingface.co/zai-org/SWE-Dev-8B} \\
    \addlinespace
    
    \multicolumn{10}{c}{\cellcolor[rgb]{0.88,0.95,1}\textbf{7B Models}} \\
    \addlinespace
    SWE-Dev-7B & Qwen2.5-Coder-7B & 7B & Dense & OpenHands & Outcome & 23.4 & \ghlink{https://github.com/THUDM/SWE-Dev} & / & \hflink{https://huggingface.co/zai-org/SWE-Dev-7B} \\
    SoRFT-7B & Qwen2.5-Coder-7B & 7B & Dense & Agentless & Outcome & 21.4 & / & / & / \\
    SEAlign-7B & Qwen2.5-Coder-7B & 7B & Dense & OpenHands & Process & 15.0 & / & / & / \\
    
    \bottomrule
    
  \end{tabular}
    \caption{A comprehensive overview of specialized models for issue resolution, categorized by parameter size. The table details each model's base architecture, the training scaffold used for rollout, the type of reward signal employed (Outcome vs. Process), and their performance results (Res. \%) on issue resolution benchmarks.}
  \label{tab:swe-specialized-models}

\end{table*}
\begin{table*}[ht!]
  \centering
  \scriptsize
  \renewcommand{\arraystretch}{1.3} 
  \setlength{\tabcolsep}{2pt}

  \begin{tabular}{
      >{\RaggedRight\bfseries}p{2.5cm}  
      c                                 
      c                                 
      >{\RaggedRight}p{2.9cm}           
      c                                 
      c                                 
      c c                               
  }
    \toprule
    \textbf{Model Name} & \textbf{Size} & \textbf{Arch.} & \textbf{Inf. Scaffold} & \textbf{Reward} & \textbf{Res.(\%)} & \textbf{Code} & \textbf{Model} \\
    \midrule
    MiMo-V2-Flash & 309B-A15B & MoE & Agentless & Outcome & 73.4 & \ghlink{https://github.com/XiaomiMiMo/MiMo-V2-Flash} & \hflink{https://huggingface.co/XiaomiMiMo/MiMo-V2-Flash} \\
    
    KAT-Coder & / & / & Claude Code & Outcome & 73.4 & / & \bloglink{https://www.modelscope.cn/models/Kwaipilot/KAT-Dev-72B-Exp} \\

    Deepseek V3.2 & 671B-A37B & MoE & Claude Code, RooCode & / & 73.1 & \ghlink{https://github.com/deepseek-ai/DeepSeek-V3.2-Exp} & \hflink{https://huggingface.co/deepseek-ai/DeepSeek-V3.2-Speciale} \\
    
    Kimi-K2-Instruct & 1T & MoE & Agentless & Outcome & 71.6 & / & \hflink{https://huggingface.co/moonshotai/Kimi-K2-Instruct} \\
    
    Qwen3-Coder & 480B-A35B & MoE & OpenHands & Outcome & 69.6 & \ghlink{https://github.com/QwenLM/Qwen3-Coder} & \hflink{https://huggingface.co/collections/Qwen/qwen3-coder} \\
    \addlinespace
    
    GLM-4.6 & 355B-A32B & MoE & OpenHands & Outcome & 68.0 & / & \hflink{https://huggingface.co/zai-org/GLM-4.6} \\
    
    gpt-oss-120b & 116.8B-A5.1B & MoE & Internal tool & Outcome & 62.0 & \ghlink{https://github.com/openai/gpt-oss} & \hflink{https://huggingface.co/openai/gpt-oss-120b} \\
    
    Minimax M2 & 230B-10B & MoE & R2E-Gym & Outcome & 61.0 & \ghlink{https://github.com/MiniMax-AI/MiniMax-M2} & \hflink{https://huggingface.co/MiniMaxAI/MiniMax-M2} \\
    \addlinespace
    
    gpt-oss-20b & 20.9B-A3.6B & MoE & Internal tool & Outcome & 60.0 & \ghlink{https://github.com/openai/gpt-oss} & \hflink{https://huggingface.co/openai/gpt-oss-20b} \\
    
    GLM-4.5-Air & 106B-A12B & MoE & OpenHands & Outcome & 57.6 & / & / \\
    
    Minimax M1-80k & 456B-A45.9B & MoE & Agentless & Outcome & 56.0 & \ghlink{https://github.com/MiniMax-AI/MiniMax-M1} & \bloglink{https://www.modelscope.cn/models/MiniMax/MiniMax-M1-80k} \\
    
    Minimax M1-40k & 456B-A45.9B & MoE & Agentless & Outcome & 55.6 & \ghlink{https://github.com/MiniMax-AI/MiniMax-M1} & \bloglink{https://www.modelscope.cn/models/MiniMax/MiniMax-M1-40k/summary} \\
    \addlinespace

    Seed1.5-Thinking & 200B-A20B & MoE & / & Outcome & 47.0 & \ghlink{https://github.com/ByteDance-Seed/Seed-Thinking-v1.5} & / \\

    Llama 4 Maverick & 400B-A17B & MoE & mini-SWE-agent & Outcome & 21.0 & \ghlink{https://github.com/meta-llama/llama-models/tree/main/models/llama4} & \hflink{https://huggingface.co/meta-llama/Llama-4-Maverick-17B-128E-Instruct} \\
    
    Llama 4 Scout & 109B-17B & MoE & mini-SWE-agent & Outcome & 9.1 & \ghlink{https://github.com/meta-llama/llama-models/tree/main/models/llama4} & \hflink{https://huggingface.co/meta-llama/Llama-4-Scout-17B-16E-Instruct} \\
    
    \bottomrule
  \end{tabular}
    \caption{Overview of general foundation models evaluated on issue resolution. The table details the specific inference scaffolds (e.g., OpenHands, Agentless) employed during the evaluation process to achieve the reported results.}
  \label{tab:general-models}

\end{table*}
Table \ref{tab:dataset-survey} provides a comprehensive survey of the rapidly evolving domain of SWE-bench related datasets. While the original SWE-bench and its refined iterations (such as SWE-bench Lite and SWE-bench Verified) set the standard for Python-based issue resolution, recent works have significantly expanded the scope of evaluation along three main axes: language diversity, modality, and scale.

\paragraph{Multilingual expansion.} A major limitation of early benchmarks was their exclusive focus on Python. To address this, datasets such as SWE-bench-java, Multi-SWE-bench, SWE-bench Multilingual, and SWE-PolyBench have been introduced. These datasets extend evaluation capabilities to a wide array of popular languages including Java, C++, Go, Rust, JavaScript, and TypeScript, encompassing thousands of repositories to test the generalization ability of agents across different syntaxes and ecosystems.

\paragraph{Multimodality.} Recognizing that modern software development often involves visual elements (particularly in frontend development), datasets like SWE-bench Multimodal, CodeV, and OmniGIRL have incorporated visual contexts. These benchmarks require agents to process not only code and text but also visual data, targeting languages like HTML, CSS, and TypeScript.

\paragraph{Scale and training resources.} To support the training of more robust agents, several large-scale datasets have been curated. SWE-Smith, SWE-Fixer, and SWE-bench-extra dramatically increase the volume of data, offering up to 115k instances (in the case of SWE-Fixer). Notably, as indicated in the "Environment" column, the field is shifting towards execution-based validation; unlike earlier static datasets, the majority of recent benchmarks now provide reproducible dockerized environments to rigorously verify agent generated patches.

\subsection{Detailed discussions on training methods}

Table \ref{tab:swe-sft-methods} provides a comprehensive overview of recent SFT-based approaches.

Table \ref{tab:swe-specialized-models} categorizes specialized models that have undergone further alignment, predominantly via Reinforcement Learning. The table sorts models by parameter size, revealing that smaller, dense models (e.g., 7B-32B) can achieve competitive performance against larger baselines when optimized with domain-specific rewards.
The trends observed in the Reward column align with the design principles detailed in Section \ref{sec:rl-methods}. While sparse outcome rewards based on test verification constitute the predominant approach, the data reveals a growing integration of process rewards. These dense feedback signals prove critical for stabilizing the training of smaller models during long-horizon tasks, thereby addressing the sparse signal challenges associated with complex repository-level debugging. Additionally, the heterogeneity observed in the Training Scaffolds category indicates a tendency to deploy RL atop established agent frameworks to enhance decision-making policies.

Finally, Table \ref{tab:general-models} lists general-purpose foundation models evaluated on issue resolution~\cite{zhan2025katcodertechnicalreport,seed2025seed15thinkingadvancingsuperbreasoning,deepseekai2025deepseekv32pushingfrontieropen,kimiteam2025kimik2openagentic,yang2025qwen3technicalreport,zeng2025glm,agarwal2025gpt,chen2025minimax,meituanlongcatteam2025introducinglongcatflashthinkingtechnicalreport,coreteam2026mimov2flashtechnicalreport}. In contrast to the specialized models above, these systems rely entirely on external inference scaffolds to bridge the gap between general reasoning and repository-level engineering. This comparison serves as a control group, highlighting the specific performance gains attributable to the SFT and RL pipelines described in Section \ref{sec:training-methods}.

\subsection{Detailed discussions on applications}
\label{sec:application-details}

The evolution of AI in software engineering is characterized by four distinct stages of increasing autonomy and architectural sophistication:

\paragraph{Stage 1: Developer augmentation.}
The initial phase is dominated by AI pair programmers, such as GitHub Copilot, which integrate directly into Integrated Development Environments (IDEs). These tools focus on real-time code completion and suggestions. Empirical evidence from enterprise adoption, such as at Accenture, indicates that these assistants can increase developer productivity by up to 55\%, leading to widespread implementation at major technology firms like Shopify.

\paragraph{Stage 2: Workflow automation.}
The subsequent level introduces AI junior developers, represented by tools like Sweep AI. These agents automate the asynchronous lifecycle of software maintenance. Unlike isolated code completions, they autonomously parse GitHub issues, analyze the relevant codebase, generate necessary modifications, and submit complete pull requests for human review, effectively handling routine engineering tasks without constant supervision.

\paragraph{Stage 3: End-to-end autonomy.}
The most advanced operational stage involves fully autonomous agents like Devin. Functioning within secure sandboxed environments equipped with a shell, editor, and browser, these agents handle complex multi-step engineering tasks ranging from planning to execution. Their industrial impact is significant; for instance, during a major code migration at the fintech company Nubank, such autonomous agents reportedly achieved a 12x improvement in engineering efficiency compared to traditional methods.

\paragraph{Stage 4: Ecosystem integration.}
The latest trend emphasizes platform interoperability and compliance. Emerging tools like Claude Code focus on embedding AI capabilities into existing enterprise workflows with rigorous security governance. Concurrently, platforms like Trae utilize the Model Context Protocol (MCP) to orchestrate multi-agent architectures, fostering an extensible ecosystem where diverse AI tools can collaborate seamlessly. However, such protocols face scalability challenges, particularly the context overhead from storing cumulative tool-invocation results. Addressing this requires protocol-level optimizations such as Anthropic’s MCP code execution to minimize the verbosity of feedback loops and maintain system efficiency.

\end{document}